\documentclass{ar-1col-S2O}
\usepackage[numbers]{natbib}
\usepackage{url}
\usepackage{amssymb}
\usepackage{amsmath}
\usepackage{amsthm}
\usepackage{mathtools}
\usepackage{comment}
\usepackage{bbm}
\newtheorem{theorem}{\bf Theorem}[section] \newtheorem{definition}{\bf Definition}[section]

\newtheorem{Algorithm}{\bf Algorithm}[section]

\usepackage{bbm}
\usepackage{bm}

\newcommand\rmd{\mathrm{d}}

\newcommand\rmf{\mathrm{f}}

\newcommand\rmp{\mathrm{p}}

\newcommand\rms{\mathrm{s}}

\newcommand\rmx{\mathrm{x}}

\newcommand\bbi{\mathbb{I}}

\newcommand\bbr{\mathbb{R}}

\newcommand\bbu{\mathbb{U}}

\newcommand\bbx{\mathbb{X}}
\newcommand\bby{\mathbb{Y}}

\usepackage{algorithm}

\jname{Annual Review of Control, Robotics, and Autonomous Systems}
\jvol{xxx}
\jyear{yyy}
\doi{10.1146/((please add article doi))}

\begin{document}

\markboth{Berberich \& Allg\"ower}{}

\title{An Overview of Systems-Theoretic Guarantees in Data-Driven Model Predictive Control
}

\author{Julian Berberich and Frank Allg\"ower
\affil{Institute for Systems Theory and Automatic Control, University of Stuttgart,
70569 Stuttgart, Germany; email: julian.berberich@ist.uni-stuttgart.de, 
frank.allgower@ist.uni-stuttgart.de}}

\begin{abstract}
The development of control methods based on data has seen a surge of interest in recent years.
When applying data-driven controllers in real-world applications, providing theoretical guarantees for the closed-loop system is of crucial importance to ensure reliable operation.
In this review, we provide an overview of data-driven model predictive control (MPC) methods for controlling unknown systems with guarantees on systems-theoretic properties such as stability, robustness, and constraint satisfaction.
The considered approaches rely on the Fundamental Lemma from behavioral theory in order to predict input-output trajectories directly from data.
We cover various setups, ranging from linear systems and noise-free data to more realistic formulations with noise and nonlinearities, and we provide an overview of different techniques to ensure guarantees for the closed-loop system.
Moreover, we discuss avenues for future research that may further improve the theoretical understanding and practical applicability of data-driven MPC.
\end{abstract}

\maketitle

\section{INTRODUCTION}\label{sec:intro}

Determining models that are both accurate and tractable, as commonly required for controller design, is challenging.
On the other hand, the increasing availability of data provides unprecedented opportunities for enhancing controllers with additional information.
System identification provides one possibility for extracting information from data by estimating a model of the underlying dynamical system~\cite{ljung1987system}, which can then be employed for control.
Obtaining theoretical guarantees for this indirect data-driven control strategy requires rigorous bounds on the estimation error, whose derivation is an active field of research~\cite{tsiamis2023statistical}.

As an alternative to indirect approaches, designing controllers directly from data without intermediate system identification has received increasing attention in the recent literature~\cite{hou2013model}.
The Fundamental Lemma by Willems et al.~\cite{willems2005note} provides a unifying framework for direct data-driven control.
It states that one persistently exciting trajectory of an unknown linear time-invariant (LTI) system can be used to parametrize all other trajectories.
This allows one to address various analysis and controller design problems, that are traditionally solved using model knowledge, based purely on data, see~\cite{markovsky2021behavioral} for a recent survey.
One of the most promising applications of the Fundamental Lemma is the design of data-driven model predictive control (MPC) schemes.
MPC is a powerful modern control technique which can handle constraints, performance criteria, and nonlinear dynamics~\cite{rawlings2020model}.
It relies on a receding-horizon principle:
At each time step, an open-loop optimal control problem is solved, the first component of the optimal input sequence is applied to the plant, and the process is repeated at the next time step using a new measurement.
While classical MPC schemes assume precise model knowledge, they can be enhanced via data using learning-based~\cite{hewing2020learning} and adaptive~\cite{adetola2011robust,tanaskovic2014adaptive,lu2021robust} approaches.

This review provides an overview of recent research on direct data-driven MPC based on the Fundamental Lemma with a focus on systems-theoretic guarantees for the closed-loop system.
It was first shown in~\cite{yang2015data,coulson2019deepc} that the Fundamental Lemma can be used to set up data-driven MPC schemes, which compute control inputs directly from data without intermediate system identification.
Since then, data-driven MPC has been successfully used in a large number of applications, including 
a quadcopter~\cite{elokda2021quadcopters}, a four-tank system~\cite{berberich2021at}, synchronous motor drives~\cite{carlet2022data}, a quadrupedal robot~\cite{fawcett2022toward}, a soft robot~\cite{mueller2022data}, green house automation~\cite{hemming2020cherry}, and traffic control~\cite{wang2023distributed,wang2023deeplcc,shang2023smoothing,rimoldi2023urban}.
Further, various successful applications have been demonstrated in the domain of energy systems, including grid-connected power converters~\cite{huang2019power,huang2021quadratic}, power system oscillation damping~\cite{huang2021decentralized}, building control~\cite{lian2021adaptive,odwyer2022automating,natale2022lessons,behrunani2023degradation}, battery charging~\cite{chen2022data}, a combined-cycle power plant~\cite{mahdavipour2022optimal}, energy networks~\cite{bilgic2022toward,schmitz2022data,yin2024data}, and a fuel cell system~\cite{schmitt2023data}.
These remarkable empirical demonstrations have motivated numerous theoretical contributions with the goal of understanding and improving data-driven MPC.
When using data-driven MPC in real-world applications, especially for complex and safety-critical systems, it is desirable to provide a priori guarantees on reliable closed-loop operation.
Mathematically, the goal is to ensure systems-theoretic properties such as stability, robustness, and constraint satisfaction for the controlled system.

In this review, we provide an overview of recent progress on systems-theoretic guarantees in data-driven MPC.
We discuss appropriate modifications that allow to prove rigorous guarantees for the closed-loop system, covering different setups including linear and nonlinear systems as well as noise-free and noisy data.
While we mainly focus on theoretical aspects of data-driven MPC, in particular closed-loop guarantees, we refer to~\cite{markovsky2021behavioral,markovsky2023data,verheijen2023handbook} for further details, e.g., on implementations.

This review is structured as follows.
In Section~\ref{sec:preliminaries}, we introduce preliminaries on the Fundamental Lemma and on model-based MPC.
Next, in Section~\ref{sec:linear}, we provide an overview of data-driven MPC for linear systems, discussing stability guarantees with noise-free data, robustifications and associated guarantees for noisy data, as well as more advanced MPC formulations.
In Section~\ref{sec:nonlinear}, we turn our attention to nonlinear data-driven MPC approaches.
These are either based on global linearity, exploiting the structure of specific system classes and knowledge of basis functions, or on local linearity, pursuing an adaptive formulation with online data updates.
Finally, Section~\ref{sec:discussion} concludes the review with a discussion and avenues for future research.

\textit{Notation:}
We denote the set of nonnegative integers by $\bbi_{\geq0}$ and integers in the interval $[a,b]$ by $\bbi_{[a,b]}$.
For a vector $\mathbf{x}$, we denote the $p$-norm by $\lVert \mathbf{x}\rVert_p$ and we define the weighted norm $\lVert \mathbf{x}\rVert_{\mathbf{P}}=\sqrt{\mathbf{x}^\top \mathbf{P}\mathbf{x}}$ for some matrix $\mathbf{P}=\mathbf{P}^\top$.
For a sequence $\{\mathbf{u}_k\}_{k=0}^{N-1}$, we define the stacked column vector $\mathbf{u}_{[k_1,k_2]}\coloneqq \begin{bmatrix}
        \mathbf{u}_{k_1}^\top&\mathbf{u}_{k_1+1}^\top&\dots&\mathbf{u}_{k_2}^\top
\end{bmatrix}^\top$
and we abbreviate $\mathbf{u}\coloneqq \mathbf{u}_{[0,N-1]}$ for the vector containing the full sequence.
Further, we define the Hankel matrix
\begin{align*}
    \mathbf{H}_L(\mathbf{u})\coloneqq\begin{bmatrix}
        \mathbf{u}_0&\mathbf{u}_1&\dots&\mathbf{u}_{N-L}\\\mathbf{u}_1&\mathbf{u}_2&\dots&\mathbf{u}_{N-L+1}\\\vdots&\ddots&\ddots&\vdots\\
        \mathbf{u}_{L-1}&\mathbf{u}_L&\dots&\mathbf{u}_{N-1}
    \end{bmatrix}\in\bbr^{mL\times(N-L+1)}.
\end{align*}

\section{Preliminaries}\label{sec:preliminaries}
In this section, we introduce the Fundamental Lemma (Section~\ref{sec:WFL}) as well as model-based MPC (Section~\ref{sec:MPC}).

\subsection{The Fundamental Lemma}\label{sec:WFL}

In the following, we consider a discrete-time linear time-invariant (LTI) system 
\begin{align}\label{eq:sys}
    \mathbf{x}_{k+1}&=\mathbf{A}\mathbf{x}_k+\mathbf{B}\mathbf{u}_k,\quad
    \mathbf{y}_k=\mathbf{C}\mathbf{x}_k+\mathbf{D}\mathbf{u}_k 
\end{align}
with state $\mathbf{x}_k\in\bbr^n$, input $\mathbf{u}_k\in\bbr^m$, output $\mathbf{y}_k\in\bbr^p$, and time $k\in\bbi_{\geq0}$.
We make the standing assumption that $(\mathbf{A},\mathbf{B})$ is controllable and $(\mathbf{A},\mathbf{C})$ is observable.
The matrices $\mathbf{A}$, $\mathbf{B}$, $\mathbf{C}$, $\mathbf{D}$, are unknown but an input-output data trajectory $\{\mathbf{u}_k^\rmd,\mathbf{y}_k^\rmd\}_{k=0}^{N-1}$ is available.
We require the corresponding input signal to be sufficiently rich in the following sense~\cite{willems2005note}.
\begin{definition}\label{def:PE}
    The signal $\{\mathbf{u}_k\}_{k=0}^{N-1}$ with $\mathbf{u}_k\in\bbr^m$ is persistently exciting (PE) of order $L$ if  $\mathbf{H}_L(\mathbf{u})$ has full row rank, i.e., $\mathrm{rank}(\mathbf{H}_L(\mathbf{u}))=mL$.
\end{definition}
The following result from~\cite{willems2005note}, commonly referred to as the Fundamental Lemma, forms the theoretical basis of data-driven MPC.
\begin{theorem}\label{thm:WFL}
    Suppose $\{\mathbf{u}_k^\rmd\}_{k=0}^{N-1}$ is PE of order $L+n$.
    Then, $\{\mathbf{\bar{u}}_k,\mathbf{\bar{y}}_k\}_{k=0}^{L-1}$ is a trajectory of the system in Equation~\ref{eq:sys} if and only if there exists $\mathbf{\alpha}\in\bbr^{N-L+1}$ such that
    \begin{align}\label{eq:thm_WFL}
        \begin{bmatrix}
            \mathbf{H}_L(\mathbf{u}^\rmd)\\\mathbf{H}_L(\mathbf{y}^\rmd)
        \end{bmatrix}\mathbf{\alpha}=\begin{bmatrix}
            \mathbf{\bar{u}}\\\mathbf{\bar{y}}
        \end{bmatrix}.
    \end{align}
\end{theorem}
The Fundamental Lemma (Theorem~\ref{thm:WFL}) provides a parametrization of all possible input-output trajectories of the system in Equation~\ref{eq:sys} based on only one input-output trajectory $\{\mathbf{u}_k^\rmd,\mathbf{y}_k^\rmd\}_{k=0}^{N-1}$.
More precisely, the image of the Hankel matrices in Equation~\ref{eq:thm_WFL} is equal to the set of all system trajectories (recall the notation $\mathbf{\bar{u}}=\mathbf{\bar{u}}_{[0,L-1]}$ and $\mathbf{\bar{y}}=\mathbf{\bar{y}}_{[0,L-1]}$).
While the result was originally formulated and proven in behavioral systems theory~\cite{willems2005note}, alternative proofs were provided more recently in the classical state-space framework~\cite{waarde2020willems,berberich2023quantitative}.
Further, the result has found increasing usage in the recent literature for developing direct data-driven analysis and control methods.
In the remainder of the review, we focus on data-driven MPC based on the Fundamental Lemma with closed-loop guarantees.
Beyond MPC, the result can be used, e.g., for data-driven simulation~\cite{markovsky2008data}, dissipativity analysis~\cite{maupong2017lyapunov,romer2019one}, and state-feedback design~\cite{persis2020formulas,berberich2020design,waarde2020informativity}, see~\cite{markovsky2021behavioral,martin2023guarantees,persis2023learning,waarde2023informativity} for recent overview articles.

\subsection{Model predictive control}\label{sec:MPC}

\begin{subequations}\label{eq:MPC}
Suppose we want to steer the system~\ref{eq:sys} to a steady-state $\mathbf{x}^\rms$ while satisfying constraints on the input $\mathbf{u}_k\in\bbu$ and the state $\mathbf{x}_k\in\bbx$ for all $k\in\bbi_{\geq0}$ with sets $\bbu\subseteq\bbr^m$, $\bbx\subseteq\bbr^n$.
In MPC, this is achieved by solving, at time $t$, the following optimal control problem.
\begin{align}\label{eq:MPC_cost}
    \underset{\mathbf{\bar{u}}(t),\mathbf{\bar{x}}(t)}{\min}&\sum_{k=0}^{L-1}
    \lVert\mathbf{\bar{x}}_k(t)-\mathbf{x}^{\rms}\rVert_{\mathbf{Q}_{\rmx}}^2+\lVert\mathbf{\bar{u}}_k(t)-\mathbf{u}^{\rms}\rVert_{\mathbf{R}}^2\\
    \label{eq:MPC_model} \text{s.t.}\>\> &\>
    \mathbf{\bar{x}}_{k+1}(t)=\mathbf{A}\mathbf{\bar{x}}_k(t)+\mathbf{B}\mathbf{\bar{u}}_k(t),\>\>k\in\bbi_{\geq0},\quad
    \mathbf{\bar{x}}_0(t)=\mathbf{x}_t,\\\label{eq:MPC_constraints}
    &\>\mathbf{\bar{u}}_k(t)\in\mathbb{U},\>\mathbf{\bar{x}}_k(t)\in\mathbb{X},\>k\in\mathbb{I}_{[0,L-1]}.
    \end{align}
    \end{subequations}
Here, $\mathbf{\bar{u}}(t)$ and $\mathbf{\bar{x}}(t)$ denote the predicted input and state trajectory at time $t$, e.g., $\mathbf{\bar{x}}_k(t)$ is the $k$-th step of the state trajectory predicted at time $t$.
According to the constraint~\ref{eq:MPC_model}, $\{\mathbf{\bar{u}}_k(t),\mathbf{\bar{x}}_k(t)\}_{k=0}^{L-1}$ is a trajectory of the system~\ref{eq:sys} which is initialized at the current state $\mathbf{x}_t$.
In the cost~\ref{eq:MPC_cost}, the distance of this trajectory to the setpoint is penalized over the prediction horizon $L$ with positive definite cost matrices $\mathbf{Q}_{\rmx}$, $\mathbf{R}$.
Further, Equation~\ref{eq:MPC_constraints} enforces the constraints on the predicted trajectory.
We denote the optimal solution of the optimization problem~\ref{eq:MPC} by $\mathbf{\bar{u}}^*(t)$, $\mathbf{\bar{x}}^*(t)$.
It can be used to synthesize a feedback controller in a receding-horizon fashion as commonly done in MPC:
At time $t$, the first component $\mathbf{\bar{u}}_0^*(t)$ of the optimal input trajectory is applied to the system and a new optimal input is computed at the next time step based on a new state measurement, see Algorithm~\ref{alg:MPC}.

When applying Algorithm~\ref{alg:MPC} without further modifications, the closed-loop system need not be stable, see~\cite{raff2006nonlinear} for an experimental example.
Guarantees on closed-loop stability and constraint satisfaction can be ensured either via a sufficiently long prediction horizon $L$~\cite{gruene2012nmpc} or by adding terminal ingredients to the optimization problem~\ref{eq:MPC}, e.g., a terminal cost and a terminal region constraint on the final state $\mathbf{\bar{x}}_L(t)$~\cite{rawlings2020model}.

\begin{algorithm}
    \begin{Algorithm}\label{alg:MPC}
    \normalfont{\textbf{Model predictive control}}\\
    \textbf{Offline:}
    Choose prediction horizon $L$, positive definite cost matrices ${\mathbf{Q}_{\rmx}},{\mathbf{R}}$, constraint sets $\mathbb{U},\mathbb{X}$, setpoint $(\mathbf{u}^\rms,\mathbf{x}^\rms)$.\\
    \textbf{Online:}
    \begin{enumerate}
    \item[1)] At time $t$, solve the optimization problem~\ref{eq:MPC}.
    \item[2)] Apply the first optimal input component $\mathbf{u}_{t}=\mathbf{\bar{u}}_0^*(t)$.
    \item[3)] Set $t=t+1$ and go back to 1).
    \end{enumerate}
    \end{Algorithm}
    \end{algorithm}

    Both the design and the implementation of standard MPC schemes as in Algorithm~\ref{alg:MPC} require model knowledge.
    For example, for solving the optimization problem~\ref{eq:MPC}, the matrices $(\mathbf{A},\mathbf{B},\mathbf{C},\mathbf{D})$ need to be available.
    In this review, we cover an alternative MPC approach which can control unknown systems only from input-output data while maintaining systems-theoretic guarantees.

\section{DATA-DRIVEN MPC FOR LINEAR SYSTEMS}\label{sec:linear}

In this section, we review data-driven MPC schemes for LTI systems with a focus on systems-theoretic guarantees for the closed loop.
We consider data-driven MPC for stabilization tasks in the ideal case of noise-free data (Section~\ref{subsec:linear_noise_free}) and in the more realistic scenario of noisy data (Section~\ref{subsec:linear_noisy}).
In Section~\ref{subsec:linear_advanced}, we discuss more advanced MPC approaches, e.g., addressing control objectives beyond stabilization.

\subsection{Data-driven MPC with noise-free data}\label{subsec:linear_noise_free}

\begin{subequations}\label{eq:DD_MPC}
    The Fundamental Lemma can be used to set up data-driven MPC schemes by replacing the commonly used state-space model (compare Section~\ref{sec:MPC}) via data-dependent Hankel matrices.
    Let us consider the control goal of tracking an input-output setpoint $(\mathbf{u}^\rms,\mathbf{y}^\rms)$ while satisfying constraints $\mathbf{u}_k\in\bbu$ and $\mathbf{y}_k\in\bby$ for all $k\in\bbi_{\geq0}$ with sets $\bbu\subseteq\bbr^m$, $\bby\subseteq\bbr^p$.
    In the following, we assume that the setpoint $(\mathbf{u}^\rms,\mathbf{y}^\rms)$ is feasible for the system dynamics~\ref{eq:sys} and we discuss possibilities for relaxing this assumption below.
    The Fundamental Lemma can be used to design an MPC scheme based only on data $\{\mathbf{u}_k^\rmd,\mathbf{y}_k^\rmd\}_{k=0}^{N-1}$ by solving, at time $t$, the following optimal control problem
\begin{align}\label{eq:DD_MPC_cost}
    \underset{\mathbf{\alpha}(t),\mathbf{\bar{u}}(t),\mathbf{\bar{y}}(t)}{\min}&\sum_{k=0}^{L-1}
    \lVert\mathbf{\bar{u}}_k(t)-\mathbf{u}^{\rms}\rVert_{\mathbf{R}}^2+\lVert\mathbf{\bar{y}}_k(t)-\mathbf{y}^{\rms}\rVert_{\mathbf{Q}}^2\\
    \label{eq:DD_MPC_hankel} \text{s.t.}\>\> &\>\begin{bmatrix}
    \mathbf{\bar{u}}(t)\\\mathbf{\bar{y}}(t)\end{bmatrix}=\begin{bmatrix}\mathbf{H}_{L+n}(\mathbf{u}^\rmd)\\\mathbf{H}_{L+n}(\mathbf{y}^\rmd)\end{bmatrix}\mathbf{\alpha}(t),
    \\\label{eq:DD_MPC_initial_conditions}
    &\>\begin{bmatrix}\mathbf{\bar{u}}_{[-n,-1]}(t)\\\mathbf{\bar{y}}_{[-n,-1]}(t)\end{bmatrix}=\begin{bmatrix}\mathbf{u}_{[t-n,t-1]}\\\mathbf{y}_{[t-n,t-1]}\end{bmatrix},\\\label{eq:DD_MPC_constraints}
    &\>\mathbf{\bar{u}}_k(t)\in\mathbb{U},\>\mathbf{\bar{y}}_k(t)\in\mathbb{Y},\>k\in\mathbb{I}_{[0,L-1]}.
    \end{align}
    \end{subequations}
    The input-output trajectory predicted at time $t$ is denoted by 
     $\mathbf{\bar{u}}(t)=\mathbf{\bar{u}}_{[-n,L-1]}(t)\in\bbu^{L+n}$,  $\mathbf{\bar{y}}(t)=\mathbf{\bar{y}}_{[-n,L-1]}(t)\in\bby^{L+n}$.
    The constraint in Equation~\ref{eq:DD_MPC_hankel} is based on the Fundamental Lemma and ensures that $\{\mathbf{\bar{u}}_k(t),\mathbf{\bar{y}}_k(t)\}_{k=-n}^{L-1}$ is a trajectory of the LTI system.
    The cost penalizes the difference of the predicted trajectory with respect to the setpoint $(\mathbf{u}^\rms,\mathbf{y}^\rms)$ for user-specified positive definite matrices ${\mathbf{Q}}$, ${\mathbf{R}}$.
    Note that the predicted trajectory is of length $L+n$.
    In the constraint~\ref{eq:DD_MPC_initial_conditions}, the first $n$ components $\{\mathbf{\bar{u}}_k(t),\mathbf{\bar{y}}_k(t)\}_{k=-n}^{-1}$ are used to initialize the prediction based on the most recent $n$ input-output measurements $\{\mathbf{u}_k,\mathbf{y}_k\}_{k=t-n}^{t-1}$.
    This initialization over $n$ time steps is required when using an input-output prediction model to ensure that the internal states of the prediction and the plant coincide, i.e., to implicitly enforce a constraint analogous to the initial condition in Equation~\ref{eq:MPC_model}.
    Instead of the system order $n$, any upper bound can be used (to be precise, an upper bound on the lag is sufficient).
    Further, Equation~\ref{eq:DD_MPC_constraints} contains the input-output constraints.
    Note that, for convex polytopic constraint sets $\bbu$, $\bby$, the optimization problem~\ref{eq:DD_MPC} is a convex quadratic program.
    We denote the optimal solution of the optimization problem~\ref{eq:DD_MPC} by $\alpha^*(t)$, $\mathbf{\bar{u}}^*(t)$, $\mathbf{\bar{y}}^*(t)$.
    It can be used to set up a data-driven MPC scheme in a receding-horizon fashion analogous to the model-based case in Section~\ref{sec:MPC}, see Algorithm~\ref{alg:DD_MPC}.

    \begin{algorithm}
        \begin{Algorithm}\label{alg:DD_MPC}
        \normalfont{\textbf{Data-driven MPC}}\\
        \textbf{Offline:}
        Choose upper bound on system order $n$, prediction horizon $L$, positive definite cost matrices ${\mathbf{Q}},{\mathbf{R}}$, constraint sets $\mathbb{U},\mathbb{Y}$, setpoint $(\mathbf{u}^\rms,\mathbf{y}^\rms)$, and generate data $\{\mathbf{u}_k^\rmd,\mathbf{y}_k^\rmd\}_{k=0}^{N-1}$.\\
        \textbf{Online:}
        \begin{enumerate}
        \item[1)] At time $t$, solve the optimization problem~\ref{eq:DD_MPC}.
        \item[2)] Apply the first optimal input component $\mathbf{u}_{t}=\mathbf{\bar{u}}_0^*(t)$.
        \item[3)] Set $t=t+1$ and go back to 1).
        \end{enumerate}
        \end{Algorithm}
        \end{algorithm}

Algorithm~\ref{alg:DD_MPC} can be used to control unknown LTI systems based only on input-output data and without explicit model knowledge.
The main difference to model-based MPC in Section~\ref{sec:MPC} is the usage of data-dependent Hankel matrices for prediction, instead of a state-space model.
If the data are PE, the optimization problem~\ref{eq:DD_MPC} returns the same optimal input as using a state-space model.
Since the Fundamental Lemma directly parametrizes input-output trajectories, Algorithm~\ref{alg:DD_MPC} inherently is an output-feedback MPC scheme.
In contrast, model-based output-feedback MPC schemes typically involve an observer.

Analogous to the model-based case, data-driven MPC may lead to an unstable closed loop and can even destabilize an open-loop stable system if $L$ is too short~\cite{berberich2021guarantees}.
Closed-loop stability of data-driven MPC can be ensured by modifying the optimization problem~\ref{eq:DD_MPC}.
The arguably simplest approach is to add terminal equality constraints, i.e., to restrict the optimal solution to be equal to the setpoint $(\mathbf{u}^\rms,\mathbf{y}^\rms)$ over $n$ steps at the end of the prediction horizon.
Mathematically, the following constraint is added to the optimization problem 
\begin{align}\label{eq:DD_MPC_TEC}
    \begin{bmatrix}\mathbf{\bar{u}}_k(t)\\\mathbf{\bar{y}}_k(t)
    \end{bmatrix}=\begin{bmatrix}
        \mathbf{u}^\rms\\\mathbf{y}^\rms
    \end{bmatrix},\>k\in\bbi_{[L-n,L-1]}.
\end{align}
This ensures that the internal state corresponding to the input-output trajectory $\mathbf{\bar{u}}(t)$, $\mathbf{\bar{y}}(t)$ is equal to the steady-state $\mathbf{x}^\rms$ corresponding to $(\mathbf{u}^\rms,\mathbf{y}^\rms)$ at time $L$.
With this modification, it can be shown under mild assumptions (e.g., PE) that, if the optimization problem is feasible at initial time $t=0$, then it is recursively feasible for all $t\in\bbi_{\geq0}$ and Algorithm~\ref{alg:DD_MPC} exponentially stabilizes $(\mathbf{u}^\rms,\mathbf{y}^\rms)$ in closed loop~\cite{berberich2021guarantees}.
While terminal equality constraints are simple to implement and to study theoretically, they have significant drawbacks in terms of robustness and the size of the region of attraction.
In particular, the latter only includes points with respect to which the optimization problem is initially feasible, i.e., from which the setpoint can be reached within $L$ time steps while satisfying the constraints.

As an alternative, one can design more general terminal ingredients, i.e., a terminal cost function and a terminal region constraint for an extended state vector consisting of sequential input-output values~\cite{berberich2021on}.
To be precise, an alternative, non-minimal state of the LTI system in Equation~\ref{eq:sys} can be obtained as 
$\mathbf{\xi}_t=\begin{bmatrix}
    \mathbf{u}_{[t-n,t-1]}\\\mathbf{y}_{[t-n,t-1]}
\end{bmatrix}$, compare~\cite{goodwin2014adaptive,koch2022provably}.
Adapting data-driven output-feedback design methods from~\cite{berberich2023combining}, it is possible to design a terminal cost function $V_f(\bar{\mathbf{\xi}}_L(t))=\lVert \bar{\mathbf{\xi}}_L(t)\rVert_{\mathbf{P}}^2$ as well as a corresponding terminal region constraint $\bar{\mathbf{\xi}}_L(t)\in\Xi_f$ for a sublevel set $\Xi_f$ of $V_f$.
Replacing the terminal equality constraint in Equation~\ref{eq:DD_MPC_TEC} by these two components leads to closed-loop stability guarantees with a significantly larger region of attraction~\cite{berberich2021on}.
However, this approach faces an important limitation:
The design of the terminal ingredients from~\cite{berberich2021on} is only applicable when the state-space realization with state $\mathbf{\xi}_t$ is controllable, which holds, e.g., if the system is single-output and the order $n$ is known exactly~\cite[Lemma 13]{berberich2021on}.
Addressing this limitation based on alternative data-driven output-feedback design approaches~\cite{waarde2024behavioral,alsalti2023notes} is an open research problem.
Beyond the above-described terminal ingredients, it can be shown that data-driven MPC without any terminal ingredients also provides desirable closed-loop guarantees if the prediction horizon is sufficiently long~\cite{bongard2023robust}.
Finally, recursive feasibility and stability can be guaranteed based on dissipativity~\cite{lazar2021dissipativity,lazar2023generalized}

All approaches discussed so far require that the setpoint $(\mathbf{u}^{\rms},\mathbf{y}^{\rms})$ is a feasible equilibrium for the unknown system, i.e., there exists $\mathbf{x}^\rms\in\bbr^n$ such that $\mathbf{x}^\rms=\mathbf{A}\mathbf{x}^\rms+\mathbf{B}\mathbf{u}^\rms,\>\>\mathbf{y}^\rms=\mathbf{C}\mathbf{x}^\rms+\mathbf{D}\mathbf{u}^\rms$.
If this is not the case, then, for example, the terminal equality constraint~\ref{eq:DD_MPC_TEC} cannot be feasible.
Feasibility of $(\mathbf{u}^{\rms},\mathbf{y}^{\rms})$ is non-trivial to verify, especially when no model is available, and therefore it can be a critical limitation for practical applications of data-driven MPC.
A powerful way to resolve this problem is to optimize over the equilibrium $(\mathbf{u}^{\rms},\mathbf{y}^{\rms})$ online and penalize its distance to the actual target setpoint in the cost.
Such an artificial equilibrium has been proposed in model-based MPC~\cite{limon2008mpc} and it leads to strong theoretical results as well as useful practical methods.
In particular, data-driven MPC with artificial equilibrium guarantees exponential stability of the unknown optimal reachable equilibrium for a given setpoint and cost function~\cite{berberich2020tracking}.
This includes, e.g., the case that only an output setpoint without corresponding input component is available.
Additional advantages include theoretical guarantees for online setpoint changes, improved robustness, and a significantly larger region of attraction compared to a standard MPC scheme with terminal equality constraints.
Further, time-varying reference trajectories can be handled as well by translating corresponding model-based MPC results~\cite{koehler2020nonlinear}.
Finally, online optimization over an artificial equilibrium plays a crucial role when using data-driven MPC to control unknown nonlinear systems, see Section~\ref{sec:nonlinear} for details.

To summarize, if the data are noise-free, then the Fundamental Lemma can be used to predict future trajectories exactly such that, under appropriate assumptions (long prediction horizon) or modifications (terminal ingredients), closed-loop stability under data-driven MPC can be guaranteed.
The main technical challenge in the analysis is that the cost function penalizes input-output values due to the employed input-output prediction model.
This only implies a positive semidefinite cost in the state, in contrast to model-based MPC which commonly assumes a positive definite cost, compare the optimization problem~\ref{eq:MPC}.
Therefore, the analysis of data-driven MPC requires additional detectability arguments~\cite{cai2008input}.

In the sidebar Subspace Predictive Control, we introduce an alternative data-driven MPC approach via an indirect perspective, i.e., using an identified model for MPC.

\begin{textbox}[h]\section{SUBSPACE PREDICTIVE CONTROL}
Subspace predictive control (SPC) involves a two-step procedure consisting of 1) identification of a multi-step predictor from data and 2) using this predictor as a model for MPC~\cite{favoreel1999spc,huang2008dynamic}.
To introduce SPC, we partition the Hankel matrices used for prediction in Equation~\ref{eq:DD_MPC_hankel} as follows 
\begin{align}
    \begin{bmatrix}\mathbf{H}_{L+n}(\mathbf{u}^\rmd)\\\mathbf{H}_{L+n}(\mathbf{y}^\rmd)\end{bmatrix}
    =
    \begin{bmatrix}
        \mathbf{H}_{n}(\mathbf{u}^\rmd_{[0,N-L-1]})\\
        \mathbf{H}_L(\mathbf{u}^\rmd_{[n,N-1]})\\
        \mathbf{H}_{n}(\mathbf{y}^\rmd_{[0,N-L-1]})\\
        \mathbf{H}_L(\mathbf{y}^\rmd_{[n,N-1]})
    \end{bmatrix}
    \eqqcolon 
    \begin{bmatrix}
        \mathbf{U}_\rmp\\\mathbf{U}_\rmf\\\mathbf{Y}_\rmp\\\mathbf{Y}_\rmf
    \end{bmatrix}.
\end{align}
Here, the data matrices $\mathbf{U}_\rmp$, $\mathbf{Y}_\rmp$ correspond to the first $n$ components of the input-output predictions, i.e., to the initial conditions in Equation~\ref{eq:DD_MPC_initial_conditions} (hence, the index 'p' for past).
On the other hand, the data matrices $\mathbf{U}_\rmf$, $\mathbf{Y}_\rmf$ correspond to future predictions (hence, the index 'f').
In SPC, the data are used to identify a multi-step predictor which predicts future output values based on a future input trajectory and initial conditions.
To be precise, the predictor $\mathbf{M}$ is determined via least-squares estimation and takes the form $\mathbf{M}=\mathbf{Y}_\rmf\begin{bmatrix}
        \mathbf{U}_\rmp\\\mathbf{U}_\rmf\\\mathbf{Y}_\rmp
    \end{bmatrix}^\dagger$,
where $\mathbf{A}^\dagger$ denotes the Moore-Penrose inverse of a matrix $\mathbf{A}$.
Based on this predictor, SPC solves the following optimal control problem at time $t$.
\begin{align}
    \min_{\mathbf{\bar{u}}(t),\mathbf{\bar{y}}(t)}\sum_{k=0}^{L-1}\lVert\mathbf{\bar{u}}_k(t)-\mathbf{u}^\rms\rVert_{\mathbf{R}}^2+\lVert\mathbf{\bar{y}}_k(t)-\mathbf{y}^\rms\rVert_{\mathbf{Q}}^2\qquad
    \text{s.t.}\>\> \>\mathbf{\bar{y}}_{[0,L-1]}(t)=\mathbf{M}\begin{bmatrix}
        \mathbf{\bar{u}}_{[-n,-1]}(t)\\
        \mathbf{\bar{u}}_{[0,L-1]}(t)\\
        \mathbf{\bar{y}}_{[-n,-1]}(t)
    \end{bmatrix}.
\end{align}
Thus, in contrast to direct data-driven MPC, SPC involves an a priori estimation step and, hence, it is an indirect data-driven MPC method.
In Section~\ref{sec:discussion}, we discuss differences between direct and indirect data-driven MPC in more detail.
\end{textbox}

\subsection{Data-driven MPC with noisy data}\label{subsec:linear_noisy}

Real-world data are rarely noise-free.
When data-driven MPC as discussed in Section~\ref{subsec:linear_noise_free} is applied in the presence of noise, there is no guarantee that the closed-loop system behaves desirably.
In the following, we describe modifications of data-driven MPC that allow for comparable theoretical guarantees also in the case of noisy data.

\begin{figure}
    \includegraphics[width=1.8in]{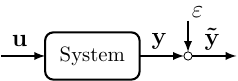}
    \caption{Setup considered for robust data-driven MPC in Section~\ref{subsec:linear_noisy}.
    The figure displays a generic system generating the data which are affected by output measurement noise $\mathbf{\tilde{y}}_k=\mathbf{y}_k+\mathbf{\varepsilon}_k$.}
    \label{fig:sys_noise}
\end{figure}

Suppose that both the offline data used for prediction via the Fundamental Lemma as well as the online data used to specify initial conditions at time $t$ are affected by output measurement noise (disturbances can be handled similarly~\cite{kloeppelt2022novel}).
To be precise, instead of the exact output values, we have access to 
\begin{align*}
    \mathbf{\tilde{y}}_k^\rmd&=\mathbf{y}_k^\rmd+\mathbf{\varepsilon}_k^\rmd,\>\> k\in\mathbb{I}_{[0,N-1]},\quad
    \mathbf{\tilde{y}}_t=\mathbf{y}_t+\mathbf{\varepsilon}_t,\>\>t\in\bbi_{\geq0},
\end{align*} 
\begin{subequations}\label{eq:DD_MPC_robust}
    compare Figure~\ref{fig:sys_noise}.
Further, the noise is bounded by some $\bar{\mathbf{\varepsilon}}>0$ in the sense that $\lVert\mathbf{\varepsilon}_k^\rmd\rVert_{\infty}\leq\bar{\mathbf{\varepsilon}}$ for $k\in\bbi_{[0,N-1]}$ and $\lVert \mathbf{\varepsilon}_t\rVert_{\infty}\leq\bar{\mathbf{\varepsilon}}$ for $t\in\bbi_{\geq0}$.
The following optimization problem provides the basis for controlling unknown linear systems based on noisy input-output data.
\begin{align}\label{eq:DD_MPC_robust_cost}
\underset{\mathbf{\alpha}(t),\mathbf{\sigma}(t),\mathbf{\bar{u}}(t),\mathbf{\bar{y}}(t)}{\min}&\sum_{k=0}^{L-1}
\lVert\mathbf{\bar{u}}_k(t)-\mathbf{u}^{\rms}\rVert_{\mathbf{R}}^2+\lVert\mathbf{\bar{y}}_k(t)-\mathbf{y}^{\rms}\rVert_{\mathbf{Q}}^2
+\lambda_{\mathbf{\alpha}}\lVert\mathbf{\alpha}(t)\rVert_2^2+\lambda_{\mathbf{\sigma}}\lVert\mathbf{\sigma}(t)\rVert_2^2\\
\label{eq:DD_MPC_robust_hankel} \text{s.t.}\>\> &\>\begin{bmatrix}
\mathbf{\bar{u}}(t)\\\mathbf{\bar{y}}(t)+\mathbf{\sigma}(t)\end{bmatrix}=\begin{bmatrix}\mathbf{H}_{L+n}(\mathbf{u}^\rmd)\\\mathbf{H}_{L+n}(\mathbf{\tilde{y}}^\rmd)\end{bmatrix}\mathbf{\alpha}(t),
\\\label{eq:DD_MPC_robust_constraints}
&\>\begin{bmatrix}\mathbf{\bar{u}}_{[-n,-1]}(t)\\\mathbf{\bar{y}}_{[-n,-1]}(t)\end{bmatrix}=\begin{bmatrix}\mathbf{u}_{[t-n,t-1]}\\\mathbf{\tilde{y}}_{[t-n,t-1]}\end{bmatrix},\\\label{eq:DD_MPC_robust_initial_conditions}
&\>\mathbf{\bar{u}}_k(t)\in\mathbb{U},\>k\in\mathbb{I}_{[0,L-1]}.
\end{align}
\end{subequations}

In contrast to data-driven MPC for noise-free data as in Section~\ref{subsec:linear_noise_free}, the exact output values are replaced by their noisy versions.
Further, the scheme contains an additional optimization variable $\mathbf{\sigma}(t)\in\bbr^{p(L+n)}$, the slack variable, which ensures that the constraint~\ref{eq:DD_MPC_robust_hankel} remains feasible despite the noise.
In order to reduce the prediction error, the slack variable is penalized in the cost with regularization parameter $\lambda_{\mathbf{\sigma}}>0$.
Moreover, the norm of the variable $\mathbf{\alpha}(t)$ is penalized with regularization parameter $\lambda_{\mathbf{\alpha}}>0$ (see Section~\ref{subsubsec:regularization} for details).
Finally, we note that the optimization problem~\ref{eq:DD_MPC_robust} does not contain output constraints, which would require a robust constraint tightening due to the noise, see~\cite{berberich2020constraints} for details.
The optimization problem~\ref{eq:DD_MPC_robust} can be used to set up a robust data-driven MPC scheme in a standard receding-horizon fashion, see Algorithm~\ref{alg:DD_MPC_robust}.

\begin{algorithm}
    \begin{Algorithm}\label{alg:DD_MPC_robust}
    \normalfont{\textbf{Robust data-driven MPC}}\\
    \textbf{Offline:}
    Choose upper bound on system order $n$, prediction horizon $L$, positive definite cost matrices ${\mathbf{Q}},{\mathbf{R}}$, regularization parameters $\lambda_{\mathbf{\alpha}},\lambda_{\mathbf{\sigma}}>0$, constraint set $\mathbb{U}$, setpoint $(\mathbf{u}^\rms,\mathbf{y}^\rms)$, and generate data $\{\mathbf{u}^\rmd_k,\mathbf{\tilde{y}}^\rmd_k\}_{k=0}^{N-1}$.\\
    \textbf{Online:}
    \begin{enumerate}
    \item[1)] At time $t$, solve the optimization problem~\ref{eq:DD_MPC_robust}.
    \item[2)] Apply the first optimal input component $\mathbf{u}_{t}=\mathbf{\bar{u}}_0^*(t)$.
    \item[3)] Set $t=t+1$ and go back to 1).
    \end{enumerate}
    \end{Algorithm}
    \end{algorithm}

In the following, we discuss several key issues in robust data-driven MPC:
closed-loop guarantees for bounded noise (Section~\ref{subsubsec:closed_loop_guarantees}), the role of the regularization of $\mathbf{\alpha}(t)$ (Section~\ref{subsubsec:regularization}), and closed-loop guarantees for stochastic noise (Section~\ref{subsubsec:stochastic_noise}).

\subsubsection{Closed-loop guarantees for bounded noise}\label{subsubsec:closed_loop_guarantees}

Similar to Section~\ref{subsec:linear_noise_free}, Algorithm~\ref{alg:DD_MPC_robust} does not guarantee closed-loop stability unless the prediction horizon is sufficiently long or terminal ingredients are added.
It was shown in~\cite{berberich2021guarantees} that stability and robustness can be ensured by adding terminal equality constraints as in Equation~\ref{eq:DD_MPC_TEC} to the optimization problem~\ref{eq:DD_MPC_robust}.
In this case, due to a technical controllability argument, proving theoretical guarantees requires the application in an $n$-step fashion~\cite{gruene2015robustness}.
This means that, in each MPC iteration, the first $n$ optimal input components $\mathbf{\bar{u}}_{[0,n-1]}^*(t)$ are applied to the system before repeating the optimization.
With these modifications, robust data-driven MPC with terminal equality constraints guarantees that the internal state $\mathbf{x}_t$ converges exponentially to a neighborhood of the steady-state $\mathbf{x}^\rms$ corresponding to $(\mathbf{u}^\rms,\mathbf{y}^\rms)$, where the size of the neighborhood depends on the noise level.
Mathematically, there exist $0<c<1$ as well as a continuous and strictly increasing function $\beta$ with $\beta(0)=0$ such that
\begin{align}
    \lVert \mathbf{x}_t-\mathbf{x}^\rms\rVert^2\leq c^t\lVert \mathbf{x}_0-\mathbf{x}^\rms\rVert^2+\beta(\bar{\mathbf{\varepsilon}}),
\end{align}
compare~\cite[Theorem 3]{berberich2021guarantees}.
For $\bar{\mathbf{\varepsilon}}=0$, the closed loop converges to the desired setpoint as in the noise-free case of Section~\ref{subsec:linear_noise_free}.
These theoretical guarantees require analogous assumptions as in the noise-free case (e.g., PE data, $n$ is an upper bound on the system order, compare Section~\ref{subsec:linear_noise_free}) and additionally that the noise level $\bar{\mathbf{\varepsilon}}$ is sufficiently small
and that the regularization parameters scale as $\lambda_{\mathbf{\alpha}}\sim\bar{\mathbf{\varepsilon}}$, $\lambda_{\mathbf{\sigma}}\sim\frac{1}{\bar{\mathbf{\varepsilon}}}$.
The latter conditions ensure that, in the noise-free case $\bar{\mathbf{\varepsilon}}=0$, robust data-driven MPC (Algorithm~\ref{alg:DD_MPC_robust}) reduces to the nominal scheme (Algorithm~\ref{alg:DD_MPC}) since $\lambda_{\mathbf{\alpha}}=0$ and $\lambda_{\mathbf{\sigma}}\to\infty$ enforces $\mathbf{\sigma}(t)=0$.
Beyond this stability result, the theoretical analysis reveals direct connections between the data quality and the closed-loop performance:
The asymptotic tracking error decreases and the region of attraction increases if the minimum singular value of the input Hankel matrix $\mathbf{H}_{L+2n}(u)$ increases, which corresponds to a quantitative PE condition~\cite{berberich2023quantitative}.
Such end-to-end insights are a valuable feature of direct data-driven control methods, which are often harder to achieve in indirect approaches.
While the original paper~\cite{berberich2021guarantees} required an additional non-convex constraint $\lVert\mathbf{\sigma}(t)\rVert_{\infty}\leq\bar{\mathbf{\varepsilon}}(1+\lVert\mathbf{\alpha}(t)\rVert_1)$ to bound the slack variable, it was later shown in~\cite{bongard2023robust} that this constraint can be dropped when $\lambda_{\mathbf{\sigma}}\sim\frac{1}{\bar{\mathbf{\varepsilon}}}$, see also~\cite[Theorem 4.1]{berberich2022phdthesis}.

As in the noise-free case, the occurrence of terminal equality constraints poses practical limitations, e.g., a small region of attraction, and alternative means of guaranteeing closed-loop stability are available.
Roughly speaking, any model-based MPC scheme which admits inherent robustness properties against disturbances can be transformed into a data-driven MPC scheme with closed-loop guarantees in the presence of noise~\cite{berberich2022inherent}.
Beyond this general principle, the literature contains various results on closed-loop guarantees in robust data-driven MPC.
For example, in the noisy data case, terminal ingredients can be designed based on robust data-driven output-feedback design~\cite{berberich2023combining}.
By combining continuity of data-driven MPC with respect to noise~\cite{berberich2022linearpart2_extended} and inherent robustness properties~\cite{berberich2022inherent,yu2014inherent}, one can show that the resulting data-driven MPC scheme admits closed-loop guarantees also in the presence of noise, compare~\cite[Section 4.5.2]{berberich2022phdthesis}.
Likewise, data-driven MPC without terminal ingredients practically exponentially stabilizes the closed-loop system, assuming that the prediction horizon $L$ is sufficiently long~\cite{bongard2023robust}.
Notably, in both cases (terminal ingredients and $L$ sufficiently long), stability and robustness can be proven even for a $1$-step MPC scheme as in Algorithm~\ref{alg:DD_MPC_robust}.
Alternatively, one can optimize over an artifical equilibrium online which provides powerful practical features (no knowledge required if $(\mathbf{u}^\rms,\mathbf{y}^\rms)$ is a feasible equilibrium, improved robustness, increased region of attraction, handling online setpoint changes) and also admits closed-loop guarantees in the presence of noise~\cite{berberich2022linearpart2_extended}.

Due to the inaccurate predictions of the Fundamental Lemma with noisy data, handling output constraints is non-trivial.
By using a bound on the prediction error from~\cite{berberich2021guarantees}, a constraint tightening can be constructed which guarantees closed-loop output constraint satisfaction~\cite{berberich2020constraints}.
The conservatism of this tightening can be reduced, e.g., via a pre-stabilizing feedback, and analogous results can be derived when the system is affected by bounded disturbances~\cite{kloeppelt2022novel}.

In summary, the Fundamental Lemma can be used to design robust data-driven MPC schemes for unknown LTI system based on noisy data.
With appropriate modifications, these schemes admit rigorous closed-loop guarantees on recursive feasibility, constraint satisfaction, and practical exponential stability.

\subsubsection{Regularization in data-driven MPC}\label{subsubsec:regularization}

While robust data-driven MPC (Algorithm~\ref{alg:DD_MPC_robust}) differs from the nominal approach (Algorithm~\ref{alg:DD_MPC}) in multiple aspects, the most crucial modification is the regularization of $\mathbf{\alpha}(t)$ in the cost.
Indeed, satisfactory practical results can often be obtained even without the slack variable $\mathbf{\sigma}(t)$.
On the contrary, the absence of the regularization of $\mathbf{\alpha}(t)$ can pose difficulties already for low noise levels or even in the complete absence of noise due to numerical inaccuracies.
Intuitively, this regularization can be understood as reducing the complexity of the employed prediction model based on the Fundamental Lemma, analogous to regularization methods in linear regression. 
In particular, due to the multiplication of $\mathbf{\alpha}(t)$ with the noisy output Hankel matrix, values $\mathbf{\alpha}(t)$ with small norm are desired to reduce the prediction error due to the noise.

Since the introduction of regularization as a heuristic in~\cite{coulson2019deepc}, numerous recent works have illuminated the role of regularization in data-driven MPC.
First, we note that regularization plays a crucial role in the closed-loop analysis of data-driven MPC, where it allows one to control the size of $\mathbf{\alpha}(t)$, which influences the prediction error~\cite{berberich2021guarantees,bongard2023robust,berberich2020constraints,berberich2022inherent,berberich2022linearpart2_extended,kloeppelt2022novel}.
Moreover, a variety of works have focused on the effect of regularization in data-driven finite-horizon optimal control, i.e., applying the full input sequence computed via the optimization problem~\ref{eq:DD_MPC_robust} in open loop.
When considering stochastic noise, the regularized optimization problem is equivalent to a min-max optimal control problem subject to distributional uncertainty, i.e., regularization serves as a robustification against noise~\cite{coulson2022distributionally}.
The regularization can also be derived through the lens of maximum likelihood estimation~\cite{yin2021predictive,yin2021maximum} or by minimizing the expected cost rather than the nominal cost~\cite{yin2023stochastic}.
Further,~\cite{breschi2023data} has proposed $\gamma$-data-driven predictive control ($\gamma$-DDPC), which decomposes the data-driven optimization problem~\ref{eq:DD_MPC_robust} into two separate steps and avoids the tuning of the regularization parameter $\lambda_{\mathbf{\alpha}}$.
The $\gamma$-DDPC scheme also allows one to construct regularization terms which systematically reduce the prediction error and can be tuned without running additional experiments~\cite{breschi2023uncertainty,breschi2023impact}.
This framework has been generalized in~\cite{chiuso2023harnessing} via the Final Control Error, which can be used to construct a data-driven MPC scheme that minimizes the actual prediction error due to the noise and can be applied without tuning regularization parameters.
Moreover, additional causality information can be used to improve performance in $\gamma$-DDPC~\cite{sader2023causality}.
Notably, equivalence between the classical regularized problem as in Algorithm~\ref{alg:DD_MPC_robust} and $\gamma$-DDPC can be established for suitable choices of the regularization parameters~\cite{klaedtke2024towards}.

While the above approaches mainly focus on LTI systems with stochastic noise, regularization can also be motivated via robustness against noise with deterministic description.
In particular, in~\cite{huang2022robust}, a data-driven optimal control problem with regularization is derived as a tractable upper bound on a min-max problem against bounded uncertainty, where different uncertainty structures lead to different regularization terms.

Regularization can also be studied through a bi-level optimization perspective, which connects direct and indirect data-driven control:
In~\cite{doerfler2023bridging}, regularization terms are derived which serve as a convex relaxation of low-rank approximation (with an $\ell_1$-norm regularization) or sequential least-squares estimation and model-based optimal control (with a regularization $\lVert \mathbf{\Pi}\mathbf{\alpha}(t)\rVert_p$ for a suitable weighting matrix $\mathbf{\Pi}$ and some $p$-norm).
The connection between different regularization approaches derived from bi-level optimization has been further investigated in~\cite{shang2023convex}.
Closely connected to these results,~\cite{klaedtke2023implicit} studies regularization via implicit predictors, which characterize the predictions caused by the constraints of the data-driven optimization problem.
Finally, regularization was used in~\cite{mattsson2023regularization,mattsson2024equivalence} to establish equivalence between direct and indirect data-driven MPC.

To conclude, regularizing the parameter $\mathbf{\alpha}(t)$ can be motivated through various reasons which are all connected to enhancing robustness.
More sophisticated regularized data-driven MPC schemes, e.g., based on maximum likelihood estimation or $\gamma$-DDPC, can further improve theoretical properties and admit practical benefits, e.g., not requiring to tune additional regularization parameters.
Finally, we note that the majority of the above works focuses on the impact of regularization on the open-loop behavior, i.e., applying the full optimal input sequence computed via the optimization problem~\ref{eq:DD_MPC_robust}.
On the other hand, studying the impact of regularization in closed loop, beyond the available robust stability results~\cite{berberich2021guarantees,bongard2023robust,berberich2020constraints,berberich2022inherent,berberich2022linearpart2_extended,kloeppelt2022novel}, and designing new regularization strategies via a closed-loop perspective provides a promising future research direction.

\subsubsection{Closed-loop guarantees for stochastic noise}\label{subsubsec:stochastic_noise}

While Section~\ref{subsubsec:closed_loop_guarantees} has focused on bounded noise, we now turn our attention to data-driven MPC for systems affected by stochastic disturbances.
One recent stream of works has employed Polynomial Chaos Expansions (PCE) for designing stochastic data-driven MPC schemes with closed-loop guarantees.
Stochastic system trajectories are random variables which can be expressed (approximately) in a polynomial basis -- the PCE basis.
By relating the images of Hankel matrices filled by three different quantities (random variables, PCE coefficients, and samples), a stochastic version of the Fundamental Lemma can be derived~\cite{pan2023stochastic}, see~\cite{faulwasser2023behavioral} for a recent tutorial and overview of behavioral data-driven control for stochastic systems.

The stochastic Fundamental Lemma can be used to construct a data-driven MPC scheme for unknown stochastic systems subject to chance constraints on the input and state.
In~\cite{pan2022towards}, it is shown that this scheme admits closed-loop guarantees on recursive feasibility and practical stability as well as closed-loop performance bound.
Here, closed-loop guarantees are ensured via terminal ingredients, i.e., a terminal cost function and a terminal region constraint, which can be constructed based only on data.
Subsequent works have extended the result in~\cite{pan2022towards}, which considers state measurements and a binary selection of the initial condition as either the measured state or a backup in case of infeasibility (the prediction from the previous time step).
In particular,~\cite{pan2022data} provides an MPC scheme with analogous guarantees for input-output measurements and~\cite{pan2023data} optimizes the initial condition, interpolating between the measured state and the backup.
Since stochastic data-driven MPC based on PCE admits an increased computational cost, a tailored multiple shooting approach for complexity reduction was derived in~\cite{ou2022data}.

As an alternative to the PCE approach, closed-loop guarantees of data-driven MPC with bounded stochastic noise can be ensured by resorting to techniques from tube-based robust MPC, i.e., computing a constraint tightening offline such that closed-loop guarantees can be established~\cite{kerz2023data}.
All previously mentioned approaches for stochastic data-driven MPC require access to past disturbance realizations and, therefore, they derive data-driven estimation procedures.
As an alternative, the recent work~\cite{teutsch2024sampling} constructs an explicit data-driven parametrization of consistent disturbance realizations from which new samples can be drawn, leading to a sampling-based stochastic data-driven MPC scheme which guarantees closed-loop properties via a robust first-step constraint.

\subsection{Data-driven MPC for more advanced control objectives}\label{subsec:linear_advanced}

Sections~\ref{subsec:linear_noise_free} and~\ref{subsec:linear_noisy} focused on data-driven MPC schemes which are designed to (robustly) stabilize an unknown LTI system.
In the following, we discuss data-driven MPC approaches with closed-loop guarantees that address more advanced setups and control objectives.

The Fundamental Lemma can be used to design distributed data-driven MPC schemes for controlling multi-agent systems with coupled dynamics, where the combination of unknown dynamics and local communication poses a key challenge.
In~\cite{allibhoy2021data,alonso2022data}, distributed data-driven MPC schemes are proposed which rely on state measurements and iterative distributed optimization and which admit closed-loop stability guarantees.
Alternatively,~\cite{koehler2022data} developed a non-iterative scheme which requires a minimum amount of communication and can cope with input-output measurements, but possibly admits increased conservatism.

Further data-driven MPC formulations in the literature include explicit MPC~\cite{breschi2023design}, which relies on an explicit solution of the underlying quadratic program via the Karush-Kuhn-Tucker conditions, and economic MPC~\cite{xie2023linear}, which addresses stage cost functions that are not positive (semi-)definite and can also handle unknown linear cost functions.
In the context of networked and cyber-physical control systems, approaches have been proposed that address resilience against denial-of-service attacks~\cite{liu2023data} as well as self-triggered~\cite{liu2023self} and event-triggered~\cite{deng2024event} MPC formulations.
Further, various contributions have been made on safe data-driven control via the Fundamental Lemma.
This includes the design of a data-driven safety filter which equips any controller with safety guarantees~\cite{bajelani2023data,bajelani2024raw}, the combination of a data-driven MPC scheme with a funnel controller~\cite{schmitz2023safe}, and enhancing data-driven MPC with control barrier certificates~\cite{khaledi2023data}.
The Fundamental Lemma as well as data-driven state-feedback design approaches~\cite{berberich2023combining,waarde2022from} have also inspired alternative data-driven MPC schemes relying on linear matrix inequalities~\cite{nguyen2023lmi,xie2024data}.
These approaches consider an infinite-horizon cost and a state-feedback parametrization of the input, and they may be less conservative than data-driven MPC when guaranteeing constraint satisfaction in the presence of noise, see~\cite{xie2024data} for a numerical comparison.
Further formulations include an extended Kalman Filter for reducing computational complexity~\cite{alpago2020extended}, data-driven MPC for linear descriptor systems~\cite{schmitz2022willems}, and data-driven control for iterative tasks~\cite{zhang2024data}.
Finally, while all the above approaches use the Fundamental Lemma for data-driven control, the dual problem of observer design has also been studied:
In~\cite{wolff2022robust}, a data-driven moving horizon estimation scheme is developed which guarantees practical exponential stability of the estimator in case of noisy data.

In conclusion, the literature contains a variety of data-driven MPC schemes for different problem setups, control objectives, and system classes.
While many of these approaches are inspired by analoguous developments in model-based MPC, the Fundamental Lemma can be used to set up MPC schemes for controlling unknown systems based only on measured data, which brings unique challenges for the theoretical analysis, in particular in the presence of uncertainty (e.g., noisy data or disturbances).

\section{DATA-DRIVEN MPC FOR NONLINEAR SYSTEMS}\label{sec:nonlinear}

While the theory of data-driven MPC is well-developed for linear systems, real-world applications to complex and safety-critical systems require more sophisticated techniques that can also cope with nonlinear systems.
In this section, we discuss data-driven MPC schemes for nonlinear systems with systems-theoretic guarantees.
As in Section~\ref{sec:linear}, the covered schemes use the Fundamental Lemma to predict future trajectories in a direct data-driven fashion.
However, the Fundamental Lemma crucially involves linearity, e.g., already the statement is based purely on linear concepts such as matrices, vector spaces, and rank conditions.
Therefore, care has to be taken when using it for nonlinear systems.

In the literature, two complementary approaches for nonlinear data-driven MPC based on the Fundamental Lemma have been developed, which we discuss in Sections~\ref{subsec:nonlinear_WFL} and~\ref{subsec:nonlinear_adaptive}, respectively.
The first class of methods rely on nonlinear versions of the Fundamental Lemma, which can be developed in a tailored fashion for specific system classes by assuming some structural knowledge, e.g., basis functions, and exploiting global linearity in higher-dimensional coordinates.
The second class follows an adaptive approach and updates the data used in the Fundamental Lemma online, exploiting that nonlinear systems are locally well-approximated by affine dynamics.
Throughout this section, we consider noise-free data but we note that the majority of the approaches can handle noisy data with modifications analogous to those discussed in Section~\ref{subsec:linear_noisy}.

\subsection{Exploiting global linearity via a nonlinear Fundamental Lemma}\label{subsec:nonlinear_WFL}

The recent literature contains various extensions of the Fundamental Lemma beyond LTI systems.
Tailored formulations have been derived for Hammerstein and Wiener systems~\cite{berberich2020trajectory}, second-order Volterra systems~\cite{rueda2020data}, linear parameter-varying (LPV) systems~\cite{verhoek2021fundamental}, affine systems~\cite{berberich2022linearpart2_extended,martinelli2022data}, nonlinear autoregressive exogenous systems~\cite{mishra2021narx}, bilinear systems~\cite{yuan2022data}, linear time-periodic systems~\cite{li2022data}, and feedback linearizable systems~\cite{alsalti2023data}.
Further, nonlinear variations of the Fundamental Lemma have been developed by resorting to multi-step predictors which are linear in basis functions~\cite{lazar2023basis}, the Koopman operator~\cite{lian2021koopman}, and kernel methods~\cite{huang2023robust,molodchyk2024exploring}.

\begin{figure}[h]
    \includegraphics[width=3.5in]{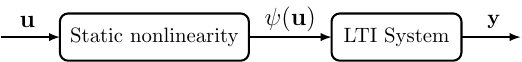}
    \caption{Block diagram illustration of a Hammerstein system, which consists of a series interconnection of a static nonlinearity and an LTI system, compare Equation~\ref{eq:sys_Hammerstein}.}
    \label{fig:sys_Hammerstein}
\end{figure}

Each of these nonlinear versions of the Fundamental Lemma allows one to design a data-driven MPC scheme for systems in the associated system class.
In the following, we show how this can be done for Hammerstein systems via the nonlinear Fundamental Lemma from~\cite{berberich2020trajectory}.
Analogous MPC schemes can be designed for other nonlinear system classes when using the corresponding version of the Fundamental Lemma.
A Hammerstein system is a series interconnection of a static nonlinearity and an LTI system, i.e.,
\begin{align}\label{eq:sys_Hammerstein}
    \mathbf{x}_{k+1}&=\mathbf{A}\mathbf{x}_k+\mathbf{B}\psi(\mathbf{u}_k),\quad
    \mathbf{y}_k=\mathbf{C}\mathbf{x}_k+\mathbf{D}\psi(\mathbf{u}_k)
\end{align}
for $k\in\bbi_{\geq0}$, compare Figure~\ref{fig:sys_Hammerstein}.
Here, $\psi:\bbr^m\to\bbr^{n_{\psi}}$ is a static nonlinear function which is assumed to be composed of known basis functions with unknown coefficients, i.e., $\psi(\mathbf{u})=\sum_{i=1}^qa_i\psi_i(\mathbf{u})$ for known functions $\psi_i:\bbr^m\to\bbr^{n_{\psi}}$ and unknown scalars $a_i\in\bbr$.

For a given data trajectory $\{\mathbf{u}_k^\rmd,\mathbf{y}_k^\rmd\}_{k=0}^{N-1}$ and under suitable PE conditions, an arbitrary input-output sequence $\{\mathbf{\bar{u}}_k,\mathbf{\bar{y}}_k\}_{k=0}^{L-1}$ is a trajectory of the system in Equation~\ref{eq:sys_Hammerstein} if and only if there exists $\alpha\in\bbr^{N-L+1}$ such that 
\begin{align}\label{eq:WFL_Hammerstein}
    \begin{bmatrix}
        \mathbf{H}_L(\mathbf{v}^\rmd)\\
        \mathbf{H}_L(\mathbf{y}^\rmd)
    \end{bmatrix}\alpha 
    =\begin{bmatrix}
        \mathbf{\bar{v}}\\\mathbf{\bar{y}}
    \end{bmatrix}
\end{align}
with auxiliary input sequences 
\begin{align}\label{eq:WFL_Hammerstein_auxiliary}
    \mathbf{v}^\rmd_k=\begin{bmatrix} \psi_1(\mathbf{u}^\rmd_k)\\\vdots\\\psi_q(\mathbf{u}^\rmd_k)\end{bmatrix},\>k\in\bbi_{[0,N-1]},\quad 
    \mathbf{\bar{v}}_k=\begin{bmatrix}
        \psi_1(\mathbf{\bar{u}}_k)\\\vdots\\
        \psi_q(\mathbf{\bar{u}}_k)
    \end{bmatrix},\>k\in\bbi_{[0,L-1]},
\end{align}
compare~\cite[Proposition 5]{berberich2020trajectory}.
This result exploits the linearity of the system which maps the auxiliary input $\mathbf{v}$, whose components involve the actual input $\mathbf{u}$ and the known basis functions $\psi_i$, to the output $\mathbf{y}$.
Since Equation~\ref{eq:WFL_Hammerstein} accurately predicts input-output trajectories of the given Hammerstein system, it can be used to set up a data-driven MPC scheme precisely as in the linear noise-free case considered in Section~\ref{subsec:linear_noise_free}.
Again, closed-loop guarantees can be established via appropriate modifications of the basic optimization problem, e.g., terminal equality constraints, terminal cost and terminal region constraints, or a sufficiently long prediction horizon.
The main limitation of this approach, beyond its restriction to Hammerstein systems and knowledge of the $\psi_i$'s, is that the resulting optimization problem to be solved at every time step is typically non-convex due to the nonlinear coupling of the decision variables $\mathbf{\bar{u}}$ and $\mathbf{\bar{v}}$ in Equation~\ref{eq:WFL_Hammerstein_auxiliary}.

An analogous data-driven MPC scheme is proposed in~\cite{alsalti2023nonlinear} for feedback linearizable systems based on the corresponding nonlinear Fundamental Lemma~\cite{alsalti2023data}.
The scheme relies on terminal equality constraints and is shown in~\cite{alsalti2022practical} to admit closed-loop guarantees on practical exponential stability, even in the presence of noisy data or inaccurate basis function approximation.
Further, the works~\cite{verhoek2021data,verhoek2023linear} address data-driven MPC for LPV systems based on the LPV Fundamental Lemma~\cite{verhoek2021fundamental}.
LPV systems provide a promising system class for data-driven control due to their ability to represent nonlinear systems while maintaining a linear structure.
In particular, the MPC scheme from~\cite{verhoek2023linear} is based on solving convex optimization problems and it includes terminal ingredients which can be designed from data via LPV state-feedback design.
Further, the MPC scheme admits closed-loop guarantees on recursive feasibility, constraint satisfaction, and exponential stability.

To summarize, nonlinear versions of the Fundamental Lemma allow to design data-driven MPC schemes with closed-loop guarantees, assuming that the underlying system indeed belongs to the associated system class.
In this case, the predictions are exact and the main challenges for the theoretical analysis are analogous to the linear case, e.g., handling positive semidefinite cost functions, designing terminal ingredients based on data, or coping with uncertainty due to noise or inexact basis function representations~\cite{alsalti2023nonlinear}.
The benefits of this theoretically sound treatment are met by possible challenges due to the assumptions on the system class.
While each of the above approaches is tailored to a specific system class, it is non-trivial to verify whether a given system indeed belongs to the assumed class.

\subsection{Exploiting local linearity via online data updates}\label{subsec:nonlinear_adaptive}

Alternatively, one can resort to an affine Fundamental Lemma and update the data used for prediction in the Hankel matrices online.
Since (smooth) nonlinear systems are locally well-approximated by affine dynamics, the resulting predictions are guaranteed to be accurate under suitable conditions.
In the following, we discuss this idea in more detail and introduce the data-driven MPC approach from~\cite{berberich2022linearpart2_extended}, which provides closed-loop guarantees when controlling unknown nonlinear systems based only on input-output data.

The control goal is to steer the output to a given setpoint $\mathbf{y}^\rms$ while satisfying input constraints $\mathbf{u}_k\in\bbu$ (guaranteeing output constraint satisfaction in this setup is an open challenge).
The considered system is assumed to be input-affine
\begin{align}\label{eq:sys_NL}
    \mathbf{x}_{k+1}&=\mathbf{f}(\mathbf{x}_k)+\mathbf{B}\mathbf{u}_k,\quad 
    \mathbf{y}_k=\mathbf{h}(\mathbf{x}_k)+\mathbf{D}\mathbf{u}_k
\end{align}
with state $\mathbf{x}_k\in\bbr^n$, input $\mathbf{u}_k\in\bbr^m$, output $\mathbf{y}_k\in\bbr^p$, and time $k\in\bbi_{\geq0}$.
The key idea is to use online data updates in order to parametrize trajectories of the affine linearization at the current state, i.e., of
\begin{align}\label{eq:sys_linearized}
    \mathbf{x}_{k+1}&=\mathbf{A}_{\mathbf{x}_t}\mathbf{x}_k+\mathbf{B}\mathbf{u}_k+\mathbf{e}_{\mathbf{x}_t},\quad 
    \mathbf{y}_k=\mathbf{C}_{\mathbf{x}_t}\mathbf{x}_k+\mathbf{D}\mathbf{u}_k+\mathbf{r}_{\mathbf{x}_t},
\end{align}
with Jacobians $\mathbf{A}_{\mathbf{x}_t}=\frac{\partial \mathbf{f}}{\partial \mathbf{x}}\Big\lvert_{\mathbf{x}_t}$, $\mathbf{C}_{\mathbf{x}_t}=\frac{\partial \mathbf{h}}{\partial \mathbf{x}}\Big\lvert_{\mathbf{x}_t}$ and remainder terms $\mathbf{e}_{\mathbf{x}_t}=\mathbf{f}(\mathbf{x}_t)-\mathbf{A}_{\mathbf{x}_t}\mathbf{x}_t$, $\mathbf{r}_{\mathbf{x}_t}=\mathbf{h}(\mathbf{x}_t)-\mathbf{C}_{\mathbf{x}_t}\mathbf{x}_t$.
Before introducing the data-driven MPC scheme, we take a slight detour to model-based MPC.
The work~\cite{berberich2022linearpart1} proposes an MPC scheme to control nonlinear systems based on the linearized dynamics in Equation~\ref{eq:sys_linearized}, i.e., an MPC scheme based on precise model knowledge.
The approach includes a terminal equality constraint with respect to an artificial equilibrium $\mathbf{x}^{\rms}(t)$ for the linearized dynamics, i.e., a vector $\mathbf{x}^{\rms}(t)$ satisfying 
\begin{align}
    \mathbf{x}^\rms(t)&=\mathbf{A}_{\mathbf{x}_t}\mathbf{x}^\rms(t)+\mathbf{B}\mathbf{u}^\rms(t)+\mathbf{e}_{\mathbf{x}_t},\quad 
    \mathbf{y}^\rms(t)=\mathbf{C}_{\mathbf{x}_t}\mathbf{x}^\rms(t)+\mathbf{D}\mathbf{u}^\rms(t)+\mathbf{r}_{\mathbf{x}_t}
\end{align}
for some equilibrium input $\mathbf{u}^\rms(t)$ and output $\mathbf{y}^\rms(t)$, compare~\cite{limon2008mpc}.
The distance of the artificial equilibrium $\mathbf{x}^{\rms}(t)$ to the actual setpoint is penalized in the cost.
If the corresponding cost weight is sufficiently small, then $\mathbf{x}^{\rms}(t)$ is encouraged to remain close to the initial state $\mathbf{x}_t$.
If $\mathbf{x}_t$ is, in turn, close to the steady-state manifold of the nonlinear system, then the full predicted trajectory $\mathbf{\bar{x}}(t)$ remains in a small region around $\mathbf{x}_t$, where the linearized dynamics in Equation~\ref{eq:sys_linearized} provide an accurate prediction.
Based on this idea, it can be shown that the MPC scheme using the linearized dynamics exponentially stabilizes the setpoint.
Notably, this guarantee does not only hold locally around the setpoint, but the region of attraction is a neighborhood of the entire steady-state manifold, which can be significantly larger.
Roughly speaking, e.g., in an autonomous driving application, successful tracking of a position can be guaranteed when starting from any initial position, under the restriction that the car does not drive too fast, i.e., it remains close to the steady-state manifold.

Let us now return to a data-driven MPC setup, where the quantities $\mathbf{f}$, $\mathbf{h}$, $\mathbf{B}$, $\mathbf{D}$ in Equation~\ref{eq:sys_NL} are unknown.
If, at time $t$, a PE input-output trajectory of the dynamics linearized at $\mathbf{x}_t$ was available, then the Fundamental Lemma for affine systems~\cite{berberich2022linearpart2_extended,martinelli2022data} would yield equivalent predictions to a state-space model.
Thus, the results in~\cite{berberich2022linearpart1} on model-based MPC could be used to guarantee closed-loop exponential stability for the corresponding data-driven MPC scheme.
However, instead of data of the linearized dynamics, in practice one only has access to measurements from the nonlinear system.
The key insight to bridge this gap lies in using, at time $t$, the most recent input-output measurements $\{\mathbf{u}_k,\mathbf{y}_k\}_{k=t-N}^{t-1}$ for prediction, i.e., the predicted trajectory $\mathbf{\bar{u}}(t)$, $\mathbf{\bar{y}}(t)$ is parametrized via
\begin{align}\label{eq:WFL_nonlinear_updates}
    \begin{bmatrix}
        \mathbf{\bar{u}}(t)\\
        \mathbf{\bar{y}}(t)
    \end{bmatrix}
    =
    \begin{bmatrix}
        \mathbf{H}_{L+n}(\mathbf{u}_{[t-N,t-1]})\\
        \mathbf{H}_{L+n}(\mathbf{y}_{[t-N,t-1]})
    \end{bmatrix}\alpha(t),\quad\sum_i\alpha_i(t)=1.
\end{align}
The first equation is analogous to the linear optimization problem~\ref{eq:DD_MPC} with the main difference of using online trajectories for prediction in the Hankel matrix rather than an offline trajectory.
The second equation ensures that the vector $\alpha(t)$ sums up to $1$, which is required due to the affine Fundamental Lemma~\cite{berberich2022linearpart2_extended,martinelli2022data} since the linearized dynamics~\ref{eq:sys_linearized} are affine.

By using an MPC scheme with artificial equilibrium as in~\cite{berberich2022linearpart1}, it can be enforced that the system does not change too rapidly in closed loop and, therefore, the trajectory $\{\mathbf{u}_k,\mathbf{y}_k\}_{k=t-N}^{t-1}$ of the nonlinear system is close to a trajectory from the dynamics linearized at $\mathbf{x}_t$.
More precisely, if a certain cost matrix in the MPC optimization problem is chosen small enough, then the closed-loop trajectory moves slowly such that Equation~\ref{eq:WFL_nonlinear_updates} provides locally accurate predictions for the nonlinear system.
Combining these ideas in a rigorous fashion, it can be shown that the setpoint is practically exponentially stable for the closed-loop system, i.e., the trajectory converges to a neighborhood around the setpoint~\cite[Theorem 2]{berberich2022linearpart2_extended}.
The size of this neighborhood depends on the distance between data samples generated initially at time $t=0,\dots,N-1$, e.g., through random exploration.
As in the model-based MPC scheme from~\cite{berberich2022linearpart1}, the guaranteed region of attraction of the data-driven MPC scheme is a neighborhood of the steady-state manifold of the underlying system.

There are several notable differences of this approach to those based on nonlinear versions of the Fundamental Lemma explained in Section~\ref{subsec:nonlinear_WFL}.
On the theoretical side, data-driven MPC with online data updates 
relies on local arguments and therefore requires different and partially less restrictive assumptions:
not restricted to systems which can be represented linearly in suitable known coordinates;
input-affine dynamics (can be enforced via an input transformation);
smooth dynamics;
properties that are also needed in model-based MPC, e.g., controllability, observability.
On the practical side, data-driven MPC with online data updates only requires input-output measurements for its implementation but no choice of basis functions for nonlinear components as in Section~\ref{subsec:nonlinear_WFL}.
Further, since the approach relies on the affine Fundamental Lemma, the resulting optimization problems are strictly convex quadratic programs, in contrast to data-driven MPC based on nonlinear basis functions.
Moreover, data-driven MPC with online data updates has been successfully applied in real-world experiments, including a four-tank system~\cite{berberich2021at} and a complex soft robot~\cite{mueller2022data}.
A related data-driven MPC scheme which also uses online data updates but does not admit closed-loop guarantees was applied to a building control experiment in~\cite{lian2021adaptive}.

On the other hand, updating the data online brings additional challenges.
In particular, the data collected from the nonlinear system in closed loop need to carry sufficient information, i.e., they need to be PE.
In practice, this can be ensured, e.g., by adding an excitation signal to the input applied to the system, by stopping the data updates once the closed-loop system has (approximately) converged, or by incentivizing PE inputs via an additional PE cost~\cite{lu2021robust}.
Addressing the problem rigorously and providing a theoretical solution is an interesting direction for future research.

\section{DISCUSSION}\label{sec:discussion}

When applying data-driven control to real-world systems, especially in complex and safety-critical applications, it is desirable to provide rigorous systems-theoretic guarantees for the closed-loop operation.
In this review, we discussed direct data-driven MPC methods based on the Fundamental Lemma which do provide such guarantees, e.g., on stability, robustness, and constraint satisfaction.
The presented schemes can control unknown systems based only on input-output data in various scenarios including linear and nonlinear systems as well as noise-free and noisy data.
As in the model-based case, closed-loop stability of data-driven MPC is not necessarily guaranteed but needs to be ensured via suitable terminal ingredients or a sufficiently long prediction horizon.
When controlling linear systems based on noise-free data, techniques from model-based MPC can be borrowed, but the technical analysis needs to additionally cope with input-output cost functions.
In the more realistic scenario of noisy data (bounded or stochastic), desirable guarantees can be provided under appropriate modifications of the MPC scheme, e.g., adding a regularization.
For nonlinear systems, we discussed two alternatives based on either global or local linearity.
The former approach exploits that systems from specific classes can be represented in linear coordinates using suitable basis functions, leading to nonlinear versions of the Fundamental Lemma that can be used for MPC.
Alternatively, nonlinear systems can be controlled by updating the data used for prediction via the Fundamental Lemma online, exploiting that (smooth) nonlinear dynamics are locally well-approximated by affine dynamics.
In guaranteeing closed-loop properties for the presented approaches, the key challenge is to combine established concepts from model-based MPC with unique challenges when using the Fundamental Lemma for prediction, in particular inaccuracies due to noise or nonlinearities.

A natural question in the context of data-driven MPC is the connection between direct approaches based on the Fundamental Lemma and indirect schemes such as SPC~\cite{doerfler2023data}.
For linear systems and noise-free data, basic formulations of SPC and data-driven MPC are equivalent since both employ exact predictions~\cite{fiedler2021relationship,lazar2022offset}.
This equivalence carries over to more advanced formulations, e.g., including regularization (compare Section~\ref{subsubsec:regularization}), and the synergy between both approaches has led to the development of new data-driven control methods~\cite{breschi2023data,lazar2023generalized,smith2024optimal}.
In the presence of inaccuracies, e.g., noisy data or nonlinearities, the predictions in direct and indirect data-driven MPC are no longer equivalent and studying performance gaps becomes relevant.
The literature contains various empirical studies, see~\cite{markovsky2021behavioral,markovsky2023data,verheijen2023handbook} and the references therein, but also theoretical results indicating that the open-loop performance of either direct or indirect approaches can be superior depending on different factors including the data length~\cite{krishnan2021on}.
Deriving formal closed-loop results on this gap, e.g., considering stability or closed-loop performance guarantees but also conservatism of constraint tightenings, is an open research problem.
To the best of the authors' knowledge, the existing SPC literature does not contain closed-loop guarantees under assumptions comparable to the direct data-driven MPC results covered in this review, in particular in the presence of noise or nonlinearities.
Deriving such guarantees may provide a meaningful step towards understanding the interplay between SPC and direct data-driven MPC. 

In terms of complexity, indirect approaches have the advantage of being less computationally demanding, whereas the number of decision variables in direct approaches grows with the data length, compare the optimization problems~\ref{eq:DD_MPC} and~\ref{eq:DD_MPC_robust}.
This has motivated contributions on complexity reduction in direct data-driven MPC, e.g., based on wavelets~\cite{sathyanarayanan2023towards}, trajectory segmentation~\cite{odwyer2023data}, singular value decomposition~\cite{zhang2023dimension}, alternative data-driven system representations~\cite{alsalti2023sample}, iterative solvers~\cite{schmitz2024fast}, and differentiable convex programming~\cite{zhou2024learning}.
Finally, another noteworthy distinction arises when designing indirect data-driven MPC schemes via identified state-space models, which admit several inherent differences to the multi-step predictor based on the Fundamental Lemma~\cite{koehler2022state}.

Data-driven MPC is a promising modern control approach with remarkable empirical performance and a solid theoretical foundation.
We conclude the review by discussing open research directions that may further enhance the theoretical understanding and reliability of data-driven MPC, thereby simplifying its application in challenging control problems.

\begin{issues}[FUTURE ISSUES]
    \begin{enumerate}
    \item Data-driven MPC for nonlinear systems:
    Data-driven methods are most useful in cases where first-principles models are hard to obtain, e.g., for complex systems with nonlinear dynamics.
    Current nonlinear data-driven MPC approaches are either tailored to specific system classes, which limits their practicality, or employ online data updates, which requires PE closed-loop trajectories.
    Developing a unifying data-driven MPC framework in the nonlinear regime, maintaining both rigorous guarantees as well as practical applicability, is an important future direction.
    
    \item Online data updates in data-driven MPC:
    The overwhelming majority of existing data-driven MPC approaches focuses on a setup where the data used for prediction via Hankel matrices are collected offline.
    On the other hand, if new data samples are obtained in closed-loop operation, it is desirable to exploit the information contained in the data to improve the performance.
    Understanding the role of data in data-driven MPC, in particular in adaptive setups where new samples enter or old data samples are deleted, is a crucial future research challenge.
    
    \item Direct vs.\ indirect data-driven MPC:
    While the interplay between direct and indirect data-driven control has received substantial attention, an ultimate assessment of their respective advantages and drawbacks requires further research.
    One promising direction is to transfer performance bounds, e.g., from~\cite{krishnan2021on}, to more general scenarios such as bounded noise, nonlinear systems, or closed-loop performance.
    
    \item Transferring open-loop results to closed-loop formulations:
    Several effects in data-driven MPC are by now well-understood in open-loop optimal control formulations, but have not been researched extensively with respect to closed-loop guarantees, e.g., regularization.
    On the other hand, when applying data-driven MPC, typically the closed-loop performance is the main object of interest.
    Thus, transferring open-loop to closed-loop results is another promising future research avenue.
    \end{enumerate}
\end{issues}

\section*{DISCLOSURE STATEMENT}
The authors are not aware of any affiliations, memberships, funding, or financial holdings that
might be perceived as affecting the objectivity of this review. 

\section*{ACKNOWLEDGMENTS}

The authors are thankful to Tim Martin for helpful comments.
F. Allg\"ower is thankful that this work was funded by the Deutsche Forschungsgemeinschaft (DFG, German Research Foundation) under Germany's Excellence Strategy - EXC 2075 - 390740016 and within grant AL 316/15-1 - 468094890.
The authors would like to thank the Stuttgart Center for Simulation Science (SimTech) for the support.

%

\bibliographystyle{ar-style3}
\bibliography{Literature}

\begin{thebibliography}{145}
\expandafter\ifx\csname natexlab\endcsname\relax\def\natexlab#1{#1}\fi

\bibitem{ljung1987system}
Ljung L. 1987.
\textit{System Identification: Theory for the User}.
Prentice-Hall, Englewood Cliffs, NJ

\bibitem{tsiamis2023statistical}
Tsiamis A, Ziemann I, Matni N, Pappas GJ. 2023.
Statistical learning theory for control: a finite-sample perspective.
\textit{IEEE Control Systems Magazine} 43(6):67--97

\bibitem{hou2013model}
Hou ZS, Wang Z. 2013.
From model-based control to data-driven control: Survey, classification and
  perspective.
\textit{Information Sciences} 235:3--35

\bibitem{willems2005note}
Willems JC, Rapisarda P, Markovsky I, {De Moor} B. 2005.
A note on persistency of excitation.
\textit{Syst. Contr. Lett.} 54:325--329

\bibitem{markovsky2021behavioral}
Markovsky I, D{\"o}rfler F. 2021.
Behavioral systems theory in data-driven analysis, signal processing, and
  control.
\textit{Annual Reviews in Control} 52:42--64

\bibitem{rawlings2020model}
Rawlings JB, Mayne DQ, Diehl MM. 2020.
\textit{Model Predictive Control: Theory, Computation, and Design}.
Nob Hill Pub.
3rd printing

\bibitem{hewing2020learning}
Hewing L, Wabersich KP, Menner M, Zeilinger MN. 2020.
Learning-based model predictive control: Toward safe learning in control.
\textit{Annual Review of Control, Robotics, and Autonomous Systems} 3:269--296

\bibitem{adetola2011robust}
Adetola V, Guay M. 2011.
Robust adaptive {MPC} for constrained uncertain nonlinear systems.
\textit{Int. J. Adaptive Control and Signal Processing} 25(2):155--167

\bibitem{tanaskovic2014adaptive}
Tanaskovic M, Fagiano L, Smith R, Morari M. 2014.
Adaptive receding horizon control for constrained {MIMO} systems.
\textit{Automatica} 50(12):3019--3029

\bibitem{lu2021robust}
Lu X, Cannon M, Koksal-Rivet D. 2021.
Robust adaptive model predictive control: performance and parameter estimation.
\textit{Int. J. Robust and Nonlinear Control} 31(18):8703--8724

\bibitem{yang2015data}
Yang H, Li S. 2015.
\textit{A data-driven predictive controller design based on reduced hankel
  matrix}.
In \textit{Proc. Asian Control Conference}, pp.  1--7

\bibitem{coulson2019deepc}
Coulson J, Lygeros J, D{\"o}rfler F. 2019.
\textit{Data-enabled predictive control: in the shallows of the {DeePC}}.
In \textit{Proc. European Control Conf. (ECC)}, pp.  307--312

\bibitem{elokda2021quadcopters}
Elokda E, Coulson J, Lygeros J, D{\"o}rfler F. 2021.
Data-enabled predictive control for quadcopters.
\textit{Int. J. Robust and Nonlinear Control} 31(18):8916--8936

\bibitem{berberich2021at}
Berberich J, K{\"o}hler J, M{\"u}ller MA, Allg{\"o}wer F. 2021{\natexlab{a}}.
Data-driven model predictive control: closed-loop guarantees and experimental
  results.
\textit{at-Automatisierungstechnik} 69(7):608--618

\bibitem{carlet2022data}
Carlet PG, Favato A, Bolognani S, D{\"o}rfler F. 2022.
Data-driven continuous-set predictive current control for synchronous motor
  drives.
\textit{IEEE Trans. Power Electronics} 37(6):6637--6646

\bibitem{fawcett2022toward}
Fawcett RT, Afsari K, Ames AD, Hamed KA. 2022.
Toward a data-driven template model for quadrupedal locomotion.
\textit{IEEE Robotics and Automation Letters} 7(3):7636--7643

\bibitem{mueller2022data}
M{\"u}ller D, Feilhauer J, Wickert J, Berberich J, Allg{\"o}wer F, Sawodny O.
  2022.
\textit{Data-driven predictive disturbance observer for quasi continuum
  manipulators}.
In \textit{Proc. 61st IEEE Conf. Decision and Control (CDC)}, pp.  1816--1822

\bibitem{hemming2020cherry}
Hemming S, {de Zwart} F, Elings E, Petropoulou A, Righini I. 2020.
Cherry tomato production in intelligent greenhouses - sensors and {AI} for
  control of climate, irrigation, crop yield, and quality.
\textit{Sensors} 20(22):6430

\bibitem{wang2023distributed}
Wang J, Lian Y, Jiang Y, Xu Q, Li K, Jones CN. 2023{\natexlab{a}}.
Distributed data-driven predictive control for cooperatively smoothing mixed
  traffic flow.
\textit{Transportation Research Part C: Emerging Technologies} 155:104274

\bibitem{wang2023deeplcc}
Wang J, Zheng Y, Li K, Xu Q. 2023{\natexlab{b}}.
{DeeP-LCC}: data-enabled predictive leading cruise control in mixed traffic
  flow.
\textit{IEEE Trans. Control Systems Technology} Doi: 10.1109/TCST.2023.3288636

\bibitem{shang2023smoothing}
Shang X, Wang J, Zheng Y. 2023.
Smoothing mixed traffic with robust data-driven predictive control for
  connected and autonomous vehicles.
\textit{arXiv:2310.00509}

\bibitem{rimoldi2023urban}
Rimoldi A, Cenedese C, Padoan A, D{\"o}rfler F, Lygeros J. 2023.
Urban traffic congestion control: a {DeePC} change.
\textit{arXiv:2311.09851}

\bibitem{huang2019power}
Huang L, Coulson J, Lygeros J, D{\"o}rfler F. 2019.
\textit{Data-enabled predictive control for grid-connected power converters}.
In \textit{Proc. 58th IEEE Conf. Decision and Control (CDC)}, pp.  8130--8135

\bibitem{huang2021quadratic}
Huang L, Zhen J, Lygeros J, D{\"o}rfler F. 2021{\natexlab{a}}.
Quadratic regularization of data-enabled predictive control: theory and
  application to power converter experiments.
\textit{IFAC-PapersOnLine} 54(7):192--197

\bibitem{huang2021decentralized}
Huang L, Coulson J, Lygeros J, D{\"o}rfler F. 2021{\natexlab{b}}.
Decentralized data-enabled predictive control for power system oscillation
  damping.
\textit{IEEE Trans. Control Systems Technology} Doi: 10.1109/TCST.2021.3088638

\bibitem{lian2021adaptive}
Lian Y, Shi J, Koch MP, Jones CN. 2021.
Adaptive robust data-driven building control via bi-level reformulation: an
  experimental result.
\textit{arXiv:2106.05740}

\bibitem{odwyer2022automating}
{O'Dwyer} E, Falugi P, Shah N, Kerrigan E. 2022.
\textit{Automating the data-driven predictive control design process for
  building thermal management}.
In \textit{Proc. 35th Int. Conf. on Efficiency, Cost, Optimization, Simulation
  and Environmental Impact of Energy Systems (ECOS)}.
Doi: 10.11581/dtu.00000267

\bibitem{natale2022lessons}
{Di Natale} L, Lian Y, Maddalena ET, Shi J, Jones CN. 2022.
\textit{Lessons learned from data-driven building control experiments:
  contrasting {Gaussian} process-based {MPC}, bilevel {DeePC}, and deep
  reinforcement learning}.
In \textit{Proc. 61st IEEE Conf. Decision and Control (CDC)}, pp.  1111--1117

\bibitem{behrunani2023degradation}
Behrunani V, Zagorowska M, {Hudoba de Badyn} M, Ricca F, Heer P, Lygeros J.
  2023.
Degradation-aware data-enabled predictive control of energy hubs.
\textit{arXiv:2307.01543}

\bibitem{chen2022data}
Chen K, Zhang X, Lin X, Zheng Y, Yin X, et~al. 2022.
Data-enabled predictive control for fast charging of {Lithium-Ion} batteries
  with constraint handling.
\textit{arXiv:2209.12862}

\bibitem{mahdavipour2022optimal}
Mahdavipour P, Wieland C, Spliethoff H. 2022.
Optimal control of combined-cycle power plants: a data-enabled predictive
  control perspective 55(13):91--96

\bibitem{bilgic2022toward}
Bilgic D, Koch A, Pan G, Faulwasser T. 2022.
Toward data-driven predictive control of multi-energy distribution systems.
\textit{Electric Power Systems Research} 212:108311

\bibitem{schmitz2022data}
Schmitz P, Engelmann A, Faulwasser T, Worthmann K. 2022.
Data-driven {MPC} of descriptor systems: a case study for power networks.
\textit{IFAC-PapersOnLine} 55(30):359--364

\bibitem{yin2024data}
Yin M, Cai H, Gattiglio A, Khayatian F, Smith RS. 2024.
Data-driven predictive control for demand side management: Theoretical and
  experimental results.
\textit{Applied Energy} 353:122101

\bibitem{schmitt2023data}
Schmitt L, Beerwerth J, Bahr M, Abel D. 2023.
Data-driven predictive control with online adaption: application to a fuel cell
  system.
\textit{IEEE Trans. Control Systems Technology} Doi: 10.1109/TCST.2023.3293790

\bibitem{markovsky2023data}
Markovsky I, Huang L, D{\"o}rfler F. 2023.
Data-driven control based on the behavioral approach: from theory to
  applications in power systems.
\textit{IEEE Control Systems}

\bibitem{verheijen2023handbook}
Verheijen PCN, Breschi V, Lazar M. 2023.
Handbook of linear data-driven predictive control: Theory, implementation and
  design.
\textit{Annual Reviews in Control} 56:100914

\bibitem{waarde2020willems}
{van Waarde} HJ, {De Persis} C, Camlibel MK, Tesi P. 2020{\natexlab{a}}.
Willems' fundamental lemma for state-space systems and its extension to
  multiple datasets.
\textit{IEEE Control Systems Lett.} 4(3):602--607

\bibitem{berberich2023quantitative}
Berberich J, Iannelli A, Padoan A, Coulson J, D{\"o}rfler F, Allg{\"o}wer F.
  2023.
\textit{A quantitative and constructive proof of {Willems' Fundamental Lemma}}.
In \textit{Proc. American Control Conf. (ACC)}, pp.  4155--4160

\bibitem{markovsky2008data}
Markovsky I, Rapisarda P. 2008.
Data-driven simulation and control.
\textit{Int. J. Control} 81(12):1946--1959

\bibitem{maupong2017lyapunov}
Maupong TM, Mayo-Maldonado JC, Rapisarda P. 2017.
On {Lyapunov} functions and data-driven dissipativity.
\textit{IFAC-PapersOnLine} 50(1):7783--7788

\bibitem{romer2019one}
Romer A, Berberich J, K{\"o}hler J, Allg{\"o}wer F. 2019.
One-shot verification of dissipativity properties from input-output data.
\textit{IEEE Control Systems Lett.} 3(3):709--714

\bibitem{persis2020formulas}
{De Persis} C, Tesi P. 2020.
Formulas for data-driven control: stabilization, optimality and robustness.
\textit{IEEE Trans. Automat. Control} 65(3):909--924

\bibitem{berberich2020design}
Berberich J, Koch A, Scherer CW, Allg{\"o}wer F. 2020{\natexlab{a}}.
\textit{Robust data-driven state-feedback design}.
In \textit{Proc. American Control Conf. (ACC)}, pp.  1532--1538

\bibitem{waarde2020informativity}
{van Waarde} HJ, Eising J, Trentelman HL, Camlibel MK. 2020{\natexlab{b}}.
Data informativity: a new perspective on data-driven analysis and control.
\textit{IEEE Trans. Automat. Control} 65(11):4753--4768

\bibitem{martin2023guarantees}
Martin T, Sch{\"o}n TB, Allg{\"o}wer F. 2023.
Guarantees for data-driven control of nonlinear systems using semidefinite
  programming: A survey.
\textit{Annual Reviews in Control} :100911

\bibitem{persis2023learning}
{De Persis} C, Tesi P. 2023.
Learning controllers for nonlinear systems from data.
\textit{Annual Reviews in Control} :100915

\bibitem{waarde2023informativity}
{van Waarde} HJ, Eising J, Camlibel MK, Trentelman HL. 2023.
The informativity approach to data-driven analysis and control.
\textit{IEEE Control Systems Magazine} 43(6):32--66

\bibitem{raff2006nonlinear}
Raff T, Huber S, Nagy ZK, Allg{\"o}wer F. 2006.
\textit{Nonlinear model predictive control of a four tank system: An
  experimental stability study}.
In \textit{Proc. IEEE Int. Conf. Control Applications (CCA)}, pp.  237--242

\bibitem{gruene2012nmpc}
Gr{\"u}ne L. 2012.
\textit{{NMPC} without terminal constraints}.
In \textit{Proc. IFAC Conf. Nonlinear Model Predictive Control}, pp.  1--13

\bibitem{berberich2021guarantees}
Berberich J, K{\"o}hler J, M{\"u}ller MA, Allg{\"o}wer F. 2021{\natexlab{b}}.
Data-driven model predictive control with stability and robustness guarantees.
\textit{IEEE Trans. Automat. Control} 66(4):1702--1717

\bibitem{berberich2021on}
Berberich J, K{\"o}hler J, M{\"u}ller MA, Allg{\"o}wer F. 2021{\natexlab{c}}.
On the design of terminal ingredients for data-driven {MPC}.
\textit{IFAC-PapersOnLine} 54(6):257--263

\bibitem{goodwin2014adaptive}
Goodwin GC, Sin KS. 2014.
\textit{Adaptive filtering prediction and control}.
Courier Corporation

\bibitem{koch2022provably}
Koch A, Berberich J, Allg{\"o}wer F. 2022.
Provably robust verification of dissipativity properties from data.
\textit{IEEE Trans. Automat. Control} 67(8):4248--4255

\bibitem{berberich2023combining}
Berberich J, Scherer CW, Allg{\"o}wer F. 2023.
Combining prior knowledge and data for robust controller design.
\textit{IEEE Trans. Automat. Control} 68(8):4618--4633

\bibitem{waarde2024behavioral}
{van Waarde} HJ, Eising J, Camlibel MK, Trentelman HL. 2024.
A behavioral approach to data-driven control with noisy input-output data.
\textit{IEEE Trans. Automat. Control} 69(2):813--827

\bibitem{alsalti2023notes}
Alsalti M, Lopez VG, M{\"u}ller MA. 2023.
Notes on data-driven output-feedback control of linear {MIMO} systems.
\textit{arXiv:2311.17484}

\bibitem{bongard2023robust}
Bongard J, Berberich J, K{\"o}hler J, Allg{\"o}wer F. 2023.
Robust stability analysis of a simple data-driven model predictive control
  approach.
\textit{IEEE Trans. Automat. Control} 68(5):2625--2637

\bibitem{lazar2021dissipativity}
Lazar M. 2021.
A dissipativity-based framework for analyzing stability of predictive
  controllers.
\textit{IFAC-PapersOnLine} 54(6):159--165

\bibitem{lazar2023generalized}
Lazar M, Verheijen PCN. 2023.
Generalized data-driven predictive control: merging subspace and {Hankel}
  predictors.
\textit{Mathematics} 11(9):2216

\bibitem{limon2008mpc}
Lim{\'o}n D, Alvarado I, Alamo T, Camacho EF. 2008.
{MPC} for tracking piecewise constant references for constrained linear
  systems.
\textit{Automatica} 44(9):2382--2387

\bibitem{berberich2020tracking}
Berberich J, K{\"o}hler J, M{\"u}ller MA, Allg{\"o}wer F. 2020{\natexlab{b}}.
Data-driven tracking {MPC} for changing setpoints.
\textit{IFAC-PapersOnLine} 53(2):6923--6930

\bibitem{koehler2020nonlinear}
K\"ohler J, M\"uller MA, Allg\"ower F. 2020.
A nonlinear tracking model predictive control scheme for dynamic target
  signals.
\textit{Automatica} 118:109030

\bibitem{cai2008input}
Cai C, Teel AR. 2008.
Input--output-to-state stability for discrete-time systems.
\textit{Automatica} 44(2):326--336

\bibitem{favoreel1999spc}
Favoreel W, Moor BD, Gevers M. 1999.
{SPC}: Subspace predictive control.
\textit{IFAC Proceedings Volumes} 32(2):4004--4009

\bibitem{huang2008dynamic}
Huang B, Kadali R. 2008.
\textit{Dynamic modeling, predictive control and performance monitoring: a
  data-driven subspace approach}.
Springer

\bibitem{kloeppelt2022novel}
Kl{\"o}ppelt C, Berberich J, Allg{\"o}wer F, M{\"u}ller MA. 2022.
A novel constraint tightening approach for robust data-driven predictive
  control.
\textit{Int. J. Robust and Nonlinear Control} Doi: 10.1002/rnc.6532

\bibitem{berberich2020constraints}
Berberich J, K{\"o}hler J, M{\"u}ller MA, Allg{\"o}wer F. 2020{\natexlab{c}}.
\textit{Robust constraint satisfaction in data-driven {MPC}}.
In \textit{Proc. 59th IEEE Conf. Decision and Control (CDC)}, pp.  1260--1267

\bibitem{gruene2015robustness}
Gr{\"u}ne L, Palma VG. 2015.
Robustness of performance and stability for multistep and updated multistep
  {MPC} schemes.
\textit{Discrete and Continuous Dynamical Systems} 35(9):4385--4414

\bibitem{berberich2022phdthesis}
Berberich J. 2022.
Stability and robustness in data-driven predictive control.
Ph.D. thesis, University of Stuttgart

\bibitem{berberich2022inherent}
Berberich J, K{\"o}hler J, M{\"u}ller MA, Allg{\"o}wer F. 2022{\natexlab{a}}.
\textit{Stability in data-driven {MPC}: an inherent robustness perspective}.
In \textit{Proc. 61st IEEE Conf. Decision and Control (CDC)}, pp.  1105--1110

\bibitem{berberich2022linearpart2_extended}
Berberich J, K{\"o}hler J, M{\"u}ller MA, Allg{\"o}wer F. 2022{\natexlab{b}}.
Linear tracking {MPC} for nonlinear systems part {II}: the data-driven case.
\textit{IEEE Trans. Automat. Control} 67(9):4406--4421Extended version on
  arXiv:2105.08567

\bibitem{yu2014inherent}
Yu S, Reble M, Chen H, Allg{\"o}wer F. 2014.
Inherent robustness properties of quasi-infinite horizon nonlinear model
  predictive control.
\textit{Automatica} 50(9):2269--2280

\bibitem{coulson2022distributionally}
Coulson J, Lygeros J, D{\"o}rfler F. 2022.
Distributionally robust chance constrained data-enabled predictive control.
\textit{IEEE Trans. Automat. Control} 67(7):3289--3304

\bibitem{yin2021predictive}
Yin M, Iannelli A, Smith RS. 2021{\natexlab{a}}.
\textit{Maximum likelihood signal matrix model for data-driven predictive
  conntrol}.
In \textit{Proc. 3rd Conf. Learning for Dynamics and Control (L4DC)}, vol. 144,
  pp.  1004--1014. PMLR

\bibitem{yin2021maximum}
Yin M, Iannelli A, Smith RS. 2021{\natexlab{b}}.
Maximum likelihood estimation in data-driven modeling and control.
\textit{IEEE Trans. Automat. Control} Doi: 10.1109/TAC.2021.3137788

\bibitem{yin2023stochastic}
Yin M, Iannelli A, Smith RS. 2023.
Stochastic data-driven predictive control: regularization, estimation, and
  constraint tightening.
\textit{arXiv:2312.02758}

\bibitem{breschi2023data}
Breschi V, Chiuso A, Formentin S. 2023.
Data-driven predictive control in a stochastic setting: a unified framework.
\textit{Automatica} 152:110961

\bibitem{breschi2023uncertainty}
Breschi V, Fabris M, Formentin S, Chiuso A. 2023{\natexlab{a}}.
Uncertainty-aware data-driven predictive control in a stochastic setting.
\textit{IFAC-PapersOnLine} 56(2):10083--10088

\bibitem{breschi2023impact}
Breschi V, Chiuso A, Fabris M, Formentin S. 2023{\natexlab{b}}.
On the impact of regularization in data-driven predictive control.
\textit{arXiv:2304.00263}

\bibitem{chiuso2023harnessing}
Chiuso A, Fabris M, Breschi V, Formentin S. 2023.
Harnessing the final control error for optimal data-driven predictive control.
\textit{arXiv:2312.14788}

\bibitem{sader2023causality}
Sader M, Wang Y, Huang D, Shang C, Huang B. 2023.
Causality-informed data-driven predictive control.
\textit{arXiv:2311.09545}

\bibitem{klaedtke2024towards}
Kl{\"a}dtke M, {Schulze Darup} M. 2024.
Towards a unifying framework for data-driven predictive control with quadratic
  regularization.
\textit{arXiv:2404.02721}

\bibitem{huang2022robust}
Huang L, Lygeros J, D{\"o}rfler F. 2022.
Robust and kernelized data-enabled predictive control for nonlinear systems.
\textit{arXiv:2206.01866}

\bibitem{doerfler2023bridging}
D{\"o}rfler F, Coulson J, Markovsky I. 2023.
Bridging direct {\&} indirect data-driven control formulations via
  regularizations and relaxations.
\textit{IEEE Trans. Automat. Control} 68(2):883--897

\bibitem{shang2023convex}
Shang X, Zheng Y. 2023.
Convex aproximations for a bi-level formulation of data-enabled predictive
  control.
\textit{arXiv:2312.15431}

\bibitem{klaedtke2023implicit}
Kl{\"a}dtke M, {Schulze Darup} M. 2023.
Implicit predictors in regularized data-driven predicive control.
\textit{IEEE Control Systems Lett.} 7:2479--2484

\bibitem{mattsson2023regularization}
Mattsson P, Sch{\"o}n TB. 2023.
On the regularization in {DeePC}.
\textit{IFAC-PapersOnLine} 56(2):625--631

\bibitem{mattsson2024equivalence}
Mattsson P, Bonassi F, Breschi V, Sch{\"o}n TB. 2024.
On the equivalence of direct and indirect data-driven predictive control
  approaches.
\textit{arXiv:2403.05860}

\bibitem{pan2023stochastic}
Pan G, Ou R, Faulwasser T. 2022{\natexlab{a}}.
On a stochastic fundamental lemma and its use for data-driven {MPC}.
\textit{IEEE Trans. Automat. Control} 68(10):5922--5937

\bibitem{faulwasser2023behavioral}
Faulwasser T, Ou R, Pan G, Schmitz P, Worthmann K. 2023.
Behavioral theory for stochastic systems? {A} data-driven journey from
  {Willems} to {Wiener} and back again.
\textit{Annual Reviews in Control} 55:92--117

\bibitem{pan2022towards}
Pan G, Ou R, Faulwasser T. 2022{\natexlab{b}}.
Towards data-driven stochastic predictive control.
\textit{arXiv:2212.10663}

\bibitem{pan2022data}
Pan G, Ou R, Faulwasser T. 2022{\natexlab{c}}.
On data-driven stochastic output-feedback predictive control.
\textit{arXiv:2211.17074}

\bibitem{pan2023data}
Pan G, Ou R, Faulwasser T. 2023.
\textit{Data-driven stochastic output-feedback predictive control: recursive
  feasibility through interpolated initial conditions}.
In \textit{Proc. 5th Conf. Learning for Dynamics and Control (L4DC)}, vol. 211,
  pp.  980--992. PMLR

\bibitem{ou2022data}
Ou R, Pan G, Faulwasser T. 2022.
Data-driven multiple shooting for stochastic optimal control.
\textit{IEEE Control Systems Lett.} 7:313--318

\bibitem{kerz2023data}
Kerz S, Teutsch J, Br{\"u}digam T, Wollherr D, Leibold M. 2023.
Data-driven tube-based stochastic predictive control.
\textit{IEEE Open Journal of Control Systems} 2:185--199

\bibitem{teutsch2024sampling}
Teutsch J, Kerz S, Wollherr D, Leibold M. 2024.
Sampling-based stochastic data-driven predictive control under data
  uncertainty.
\textit{arXiv:2402.00681}

\bibitem{allibhoy2021data}
Allibhoy A, Cort{\'e}s J. 2021.
Data-based receding horizon control of linear network systems.
\textit{IEEE Control Systems Lett.} 5(4):1207--1212

\bibitem{alonso2022data}
Alonso CA, Yang F, Matni N. 2022.
Data-driven distributed and localized model predictive control.
\textit{IEEE Open Journal of Control Systems} 1:29--40

\bibitem{koehler2022data}
K{\"o}hler M, Berberich J, M{\"u}ller MA, Allg{\"o}wer F. 2022{\natexlab{a}}.
Data-driven distributed {MPC} of dynamically coupled linear systems.
\textit{IFAC-PapersOnLine} 55(30):365--370

\bibitem{breschi2023design}
Breschi V, Sassella A, Formentin S. 2023.
On the design of regularized explicit predictive controllers from input-output
  data.
\textit{IEEE Trans. Automat. Control} 68(8):4977--4983

\bibitem{xie2023linear}
Xie Y, Berberich J, Allg{\"o}wer F. 2023.
Linear data-driven economic {MPC} with generalized terminal constraint.
\textit{IFAC-PapersOnLine} 56(2):5512--5517

\bibitem{liu2023data}
Liu W, Sun J, Wang G, Bullo F, Chen J. 2023{\natexlab{a}}.
Data-driven resilient predictive control under denial-of-service.
\textit{IEEE Trans. Automat. Control} 68(8):4722--4737

\bibitem{liu2023self}
Liu W, Sun J, Wang G, Bullo F, Chen J. 2023{\natexlab{b}}.
Data-driven self-triggered control via trajectory prediction.
\textit{IEEE Trans. Automat. Control} 68(11):6951--6958

\bibitem{deng2024event}
Deng L, Shu Z, Chen T. 2024.
Event-triggered robust {MPC} with terminal inequality constraints: a
  data-driven approach.
\textit{IEEE Trans. Automat. Control} Doi: 10.1109/TAC.2024.3357417

\bibitem{bajelani2023data}
Bajelani M, {van Heusden} K. 2023.
Data-driven safety filter: An input-output perspective.
\textit{arXiv:2309.00189}

\bibitem{bajelani2024raw}
Bajelani M, {van Heusden} K. 2024.
From raw data to safety: reducing conservatism by set expansion.
\textit{arXiv:2403.15883}

\bibitem{schmitz2023safe}
Schmitz P, Lanza L, Worthmann K. 2023.
\textit{Safe data-driven reference tracking with prescribed performance}.
In \textit{Proc. 27th Int. Conf. on System Theory, Control and Computing
  (ICSTCC)}, pp.  454--460

\bibitem{khaledi2023data}
Khaledi M, Tooranjipour P, Kiumarsi B. 2023.
Data-driven safety-certified predictive control for linear systems.
\textit{IEEE Control Systems Lett.} 7:3687--3692

\bibitem{waarde2022from}
{van Waarde} HJ, {Camlibel} MK, Mesbahi M. 2022.
From noisy data to feedback controllers: non-conservative design via a matrix
  {S}-lemma.
\textit{IEEE Trans. Automat. Control} 67(1):162--175

\bibitem{nguyen2023lmi}
Nguyen HH, Friedel M, Findeisen R. 2023.
{LMI}-based data-driven robust model predictive control.
\textit{IFAC-PapersOnLine} 56(2):4783--4788

\bibitem{xie2024data}
Xie Y, Berberich J, Allg{\"o}wer F. 2024.
Data-driven min-max {MPC} for linear systems: robustness and adaptation.
\textit{arXiv:2404.19096}

\bibitem{alpago2020extended}
Alpago D, D{\"o}rfler F, Lygeros J. 2020.
An extended {Kalman} filter for data-enabled predictive control.
\textit{IEEE Control Systems Lett.} 4(4):994--999

\bibitem{schmitz2022willems}
Schmitz P, Faulwasser T, Worthmann K. 2022.
Willems' fundamental lemma for linear descriptor systems and its use for
  data-driven output-feedback {MPC}.
\textit{IEEE Control Systems Lett.} 6:2443--2448

\bibitem{zhang2024data}
Zhang K, Zuliani R, Balta EC, Lygeros J. 2024.
Data-enabled predictive iterative control.
\textit{IEEE Control Systems Lett.} Doi: 10.1109/LCSYS.2024.3408073

\bibitem{wolff2022robust}
Wolff TM, Lopez VG, M{\"u}ller MA. 2022.
Robust data-driven moving horizon estimation for linear discrete-time systems.
\textit{arXiv:2210.09017}

\bibitem{berberich2020trajectory}
Berberich J, Allg\"ower F. 2020.
\textit{A trajectory-based framework for data-driven system analysis and
  control}.
In \textit{Proc. European Control Conf. (ECC)}, pp.  1365--1370

\bibitem{rueda2020data}
Rueda-Escobedo JG, Schiffer J. 2020.
\textit{Data-driven internal model control of second-order discrete {Volterra}
  systems}.
In \textit{Proc. 59th IEEE Conf. Decision and Control (CDC)}, pp.  4572--4579

\bibitem{verhoek2021fundamental}
Verhoek C, T{\'o}th R, Haesart S, Koch A. 2021{\natexlab{a}}.
\textit{Fundamental Lemma for data-driven analysis of linear parameter-varying
  systems}.
In \textit{Proc. 60th IEEE Conf. Decision and Control (CDC)}, pp.  5040--5046

\bibitem{martinelli2022data}
Martinelli A, Gargiani M, Draskovic M, Lygeros J. 2022.
Data-driven optimal control of affine systems: a linear programming
  perspective.
\textit{arXiv:2203.12044}

\bibitem{mishra2021narx}
Mishra VK, Markovsky I, Fazzi A, Dreesen P. 2021.
\textit{Data-driven simulation for {NARX} systems}.
In \textit{Proc. European Signal Processing Conference}, pp.  1055--1059

\bibitem{yuan2022data}
Yuan Z, Cort{\'e}s J. 2022.
Data-driven optimal control of bilinear systems.
\textit{IEEE Control Systems Lett.} 6:2479--2484

\bibitem{li2022data}
Li R, Simpson-Porco JW, Smith SL. 2022.
\textit{Data-driven model predictive control for linear time-periodic systems}.
In \textit{Proc. 61st IEEE Conf. Decision and Control}, pp.  3661--3668

\bibitem{alsalti2023data}
Alsalti M, Lopez VG, Berberich J, Allg{\"o}wer F, M{\"u}ller MA.
  2023{\natexlab{a}}.
Data-based control of feedback linearizable systems.
\textit{IEEE Trans. Automat. Control} 68(11):7014--7021

\bibitem{lazar2023basis}
Lazar M. 2023.
Basis functions nonlinear data-enabled predictive control: consistent and
  computationally efficient formulations.
\textit{arXiv:2311.05360}

\bibitem{lian2021koopman}
Lian Y, Wang R, Jones CN. 2021.
Koopman based data-driven predictive control.
\textit{arXiv:2102.05122}

\bibitem{huang2023robust}
Huang L, Zhen J, Lygeros J, D{\"o}rfler F. 2023.
Robust data-enabled predictive control: tractable formulations and performance
  guarantees.
\textit{IEEE Trans. Automat. Control} 68(5):3163--3170

\bibitem{molodchyk2024exploring}
Molodchyk O, Faulwasser T. 2024.
Exploring the links between the {Fundamental Lemma} and kernel regression.
\textit{arXiv:2403.05368}

\bibitem{alsalti2023nonlinear}
Alsalti M, Lopez VG, Berberich J, Allg{\"o}wer F, M\"uller MA.
  2023{\natexlab{b}}.
Data-driven nonlinear predictive control for feedback linearizable systems.
\textit{IFAC-PapersOnLine} 56(2):617--624

\bibitem{alsalti2022practical}
Alsalti M, Lopez VG, Berberich J, Allg{\"o}wer F, M{\"u}ller MA. 2022.
Practical exponential stability of a robust data-driven nonlinear predictive
  control scheme.
\textit{arXiv:2204.01150} Suppl.\ mat.

\bibitem{verhoek2021data}
Verhoek C, Abbas HS, T{\'o}th R, Haesart S. 2021{\natexlab{b}}.
Data-driven predictive control for linear parameter-varying systems.
\textit{IFAC-PapersOnLine} 54(8):101--108

\bibitem{verhoek2023linear}
Verhoek C, Berberich J, Haesaert S, T{\'o}th R, Abbas HS. 2023.
A linear parameter-varying approach to data predictive control.
\textit{arXiv:2311.07140}

\bibitem{berberich2022linearpart1}
Berberich J, K{\"o}hler J, M{\"u}ller MA, Allg{\"o}wer F. 2022{\natexlab{c}}.
Linear tracking {MPC} for nonlinear systems part {I}: the model-based case.
\textit{IEEE Trans. Automat. Control} 67(9):4390--4405

\bibitem{doerfler2023data}
D{\"o}rfler F. 2023.
Data-driven control: part two of two: hot take: why not go with models?
\textit{IEEE Control Systems Magazine} 43(6):27--31

\bibitem{fiedler2021relationship}
Fiedler F, Lucia S. 2021.
\textit{On the relationship between data-enabled predictive control and
  subspace predictive control}.
In \textit{Proc. European Control Conf. (ECC)}, pp.  222--229

\bibitem{lazar2022offset}
Lazar M, Verheijen PCN. 2022.
\textit{Offset-free data-driven predictive control}.
In \textit{Proc. 61st IEEE Conf. Decision and Control (CDC)}, pp.  1099--1104

\bibitem{smith2024optimal}
Smith RS, Abdalmoaty M, Yin M. 2024.
Optimal data-driven prediction and predictive control using signal matrix
  models.
\textit{arXiv:2403.15329}

\bibitem{krishnan2021on}
Krishnan V, Pasqualetti F. 2021.
\textit{On direct vs indirect data-driven predictive control}.
In \textit{Proc. 60th IEEE Conf. Decision and Control (CDC)}, pp.  736--741

\bibitem{sathyanarayanan2023towards}
Sathyanarayanan KK, Pan G, Faulwasser T. 2023.
Towards data-driven predictive control using wavelets.
\textit{IFAC-PapersOnLine} 56(2):632--637

\bibitem{odwyer2023data}
{O'Dwyer} E, Kerrigan EC, Falugi P, Zagorowska M, Shah N. 2023.
Data-driven predictive control with improved performance using segmented
  trajectories.
\textit{IEEE Trans. Control Systems Technology} 31(3):1355--1365

\bibitem{zhang2023dimension}
Zhang K, Zheng Y, Shang C, Li Z. 2023.
Dimension reduction for efficient data-enabled predictive control.
\textit{IEEE Control Systems Lett.} 7:3277--3282

\bibitem{alsalti2023sample}
Alsalti M, Barkey M, Lopez VG, M{\"u}ller MA. 2023{\natexlab{c}}.
Sample- and computationally efficient data-driven predictive control.
\textit{arXiv:2309.11238}

\bibitem{schmitz2024fast}
Schmitz P, Schaller M, Voigt M, Worthmann K. 2024.
Fast and memory-efficient optimization for large-scale data-driven predictive
  control.
\textit{arXiv:2402.13090}

\bibitem{zhou2024learning}
Zhou Y, Lu Y, Li Z, Yan J, Mo Y. 2024.
Learning-based efficient approximation of data-enabled predictive control.
\textit{arXiv:2404.16727}

\bibitem{koehler2022state}
K{\"o}hler J, Wabersich KP, Berberich J, Zeilinger MN. 2022{\natexlab{b}}.
\textit{State space models vs. multi-step predictors in predictive control: Are
  state space models complicating safe data-driven designs?}
In \textit{Proc. 61st IEEE Conf. Decision and Control (CDC)}, pp.  491--498

\end{thebibliography}

\end{document}